\documentclass[12pt]{iopart}

\usepackage{psfig}
\usepackage{epsfig}
\usepackage{bm}
\usepackage[nonstop]{feynmf}
\usepackage[T1]{fontenc}
\usepackage{ae,aecompl}
\usepackage{amssymb}
\usepackage{mathrsfs}

\newcommand{\tgg}{\mathscr{G}}

\newcommand{\tgf}{\mathscr{F}}

\newcommand{\cpsi}[2]{\hat{\psi}^\dag_{#1}({#2})}
\newcommand{\dpsi}[2]{\hat{\psi}_{#1}({#2})}
\newcommand{\up}{\uparrow}
\newcommand{\down}{\downarrow}
\newcommand{\vr}{{\bf r}}

\begin{document}
\begin{fmffile}{fmf}

\title{Quasi-particle properties of trapped Fermi gases}
\author{Luca Giorgetti$^1$, Luciano Viverit$^{1}$, Giorgio Gori$^{1}$, Francisco Barranco$^2$, Enrico Vigezzi$^{1}$,
and Ricardo A. Broglia$^{1,3}$}
\address{$^1$Dipartimento di Fisica, Universit\`a di Milano and INFN Sezione di Milano, via Celoria 16, 20133 Milano, Italy}
\address{$^2$Departamento de Fisica Aplicada III, Escuela Superior de Ingenieros,
camino de los Descubrimientos s/n, 41029 Sevilla, Spain}
\address{$^3$The Niels Bohr Institute, University of Copenhagen,
Blegdamsvej 17, 2100 Copenhagen, Denmark}

\begin{abstract}
We develop a consistent formalism in order to explore the effects of
density and spin fluctuations on the quasi-particle properties and
on the pairing critical temperature of a trapped
Fermi gas on the attractive side of a Feshbach resonance. We first
analyze the quasi-particle properties of a gas due to interactions
far from resonance (effective mass and lifetime, quasi-particle
strength and effective interaction) for the two cases of a
spherically symmetric harmonic trap and of a spherically symmetric
infinite potential well. We then explore the effect of each of these
quantities on $T_c$ and point out the important role played by the
discrete level structure.
\end{abstract}


\maketitle

\section{Introduction}

In the last few years the progress in cooling and manipulation of
atomic Fermi gases has been impressive
\cite{EXPTS1,EXPTS2,EXPTS3,EXPTS4,EXPTS5}. The most notable feature
of atomic gases is the presence of scattering resonances (Feshbach
resonances), which can be addressed using an external magnetic
field. The resonance is induced in the scattering between two atoms
in different internal states, typically hyperfine states, and
results in the divergence of the two-body $s$-wave scattering length
$a_F$.  Feshbach resonances allow the experimental study of a Fermi
gas at various interaction regimes. The crucial feature is that, by
varying the value of the scattering length, one can explore
different kinds of fermionic superfuidity, ranging from the
weak-coupling BCS superfluid, for $a_F$ small and negative, to the
strong-coupling unitarity regime, $a_F\to \infty$, and finally to
the Bose-Einstein condensation regime, $a_F>0$, where the fermions
couple pair-wise to form bosonic molecular bound states. In fact, it
has been pointed out some years ago that BCS and BEC are two aspects
of one and the same phenomenon, and the basic theory which connects
the two was then developed \cite{LEGGETT,NOZIERESSCHMITT,SA}.
Recently, this theory came again to the attention of the scientific
community in the context of cold atomic Fermi gases
\cite{THEORY1,THEORY2,THEORY3}. One can show that BCS and BEC appear
at the opposite sides of the resonance if one keeps track of the
mean-field normal and anomalous pair correlation functions. This
essentially corresponds to extending mean-field BCS theory to the
whole crossover to BEC. Although the resulting theory is remarkably
elegant, it is only qualitatively correct. The absence of important
ingredients from the theory as the fluctuating parts of the pair
correlation functions and the higher order correlations leads to
various flaws in the quantitative predictions. Keeping the full
correlations is of course impracticable unless one resorts to
quantum Monte Carlo calculations \cite{CARLSON1,CARLSON2,GIORGINI}.
To pinpoint and understand the important physics contained in the
higher order correlations, however, it is also useful to apply
approximate methods. This is the line we follow in the present work.

We are interested in improving our understanding of the properties of
a Fermi gas as the interaction strength increases originating from the
weak-coupling BCS regime. In particular the key quantity we calculate
is the critical temperature for pairing $T_c$. Moving towards the
resonance, higher order correlations and fluctuations over the
mean-field become important. One may expect the emission and
reabsorption of density and spin fluctuations to play an important
role. There are very close analogies between a gas in these conditions
and a Fermi liquid. One can build a theory along similar lines
introducing the concept of quasi-particles displaying an effective
mass, a finite lifetime and an effective interaction due to coupling
with fluctuations.  The current experiments with Fermi gases have one
more particularly interesting aspect in that the atoms are usually
confined in harmonic potentials. The finite size plays an important
role when the number of particles is not large, as is also the case in
optical lattices with few atoms per lattice site, when tunneling
between neighboring sites is suppressed and atoms are subjected to an
effective harmonic confinement. Both these situations are of interest
and have close relations with the physics of finite nuclei and atomic
clusters (cf. e. g. \cite{BRINK,GUNNARSSON,COLO}). In this article we
first develop a formalism appropriate to finite size systems confined
by a general, isotropic potential. Subsequently we apply the formalism
for the two specific cases of an infinite square well, which we
compare with an homogeneous system, and of a spherically symmetric
harmonic confinement (cf. also \cite{PREPRINT}).  The results
illustrate the central role the discrete level structure plays in
determining the properties of these systems.  Finally we relate our
approach to other calculations for $T_c$ (or equivalently for the
$T=0$ pairing gap) made in different fields.

The paper in organized as follows. In Sec. \ref{sec:theory} we
present the theoretical framework. We show the approximations used
which lead to the emergence of the renormalized quasi-particle
properties, namely the normal self-energy (Sec. \ref{sec:normal}),
the anomalous self-energy and the induced interaction (Sec.
\ref{sec:anomalous}), and the one pole approximation (Sec.
\ref{sec:onepole}). Finally we derive the equations used to
calculate $T_c$ (Sec. \ref{sec:gap}), including the multipolar
expansion appropriate for an isotropic system (Sec.
\ref{sec:multi}). In Sec. \ref{sec:results}, which is divided into
two main subsections, we report the results of the numerical
calculations. Sec. \ref{sec:resunif} contains the results for the
infinite square well. We discuss the quasi-particle properties and
compare the results obtained with those of an infinite system.  We
also present the calculated values for $T_c$.  In Sec.
\ref{sec:restrap} the same subjects and corresponding results for a
harmonically trapped system are presented. Section
\ref{sec:conclusions} contains the conclusions. In
\ref{appcorr} details concerning the correlation functions and the
linear response of the system for both the uniform and harmonically
trapped case are collected.

Before proceeding a remark on notation. In a uniform system the many-body
interaction strength is parametrized by $\lambda=gN(0)$, with $g$ being the
coupling constant (related to the scattering length $a_F$) and
$N(0)=mk_F/2\pi^2\hbar^2$ the density of states at the Fermi surface for a
single spin orientation.  For such system the Fermi momentum $k_F$ is related
to the uniform density $n$ of the gas by $(3\pi^2)^{1/3} n^{1/3}$. A similar
definition can be introduced for a trapped gas using in place of $n$ the
density at the center of the interacting cloud.

\section{Theoretical tools}

\label{sec:theory}

We consider a Fermi gas of $N$ atoms in two internal states
with equal numbers ($N/2$) described by the Hamiltonian
\begin{eqnarray}
  \nonumber
  \hat{H}&=&\sum_{\sigma=\up,\down}\int d^3 r\;
  \hat{\psi}^{\dagger}_{\sigma}({\bf r})(H_0 -\mu)\hat{\psi}_{\sigma}({\bf
    r})\\ &+&g\int d^3 r\;
  \hat{\psi}^{\dagger}_{\up}({\bf r})
  \hat{\psi}^{\dagger}_{\down}({\bf r})
  \hat{\psi}_{\down}({\bf r})
  \hat{\psi}_{\up}({\bf r}).
  \label{Eq:Hamil}
\end{eqnarray}
Here $H_0=-\hbar^2\nabla^2/2m_a+V_{ext}(r)$, with $m_a$ being the
atomic mass, and $\uparrow$ and $\downarrow$ conventionally label the
two atomic internal states. The only assumptions made at this level on
the external potential are that it is isotropic and that it acts in
the same way on both atomic states. These assumptions imply that the
following relation holds for the chemical potentials
$\mu_\up=\mu_\down=\mu$. Finally, we are here interested in the BCS
side of the resonance where $g<0$.

The properties of the system at any given temperature $T$ can be
described introducing the normal and anomalous propagators
$\tgg_{\up\up}(\vr,\vr',\omega_n)$ and
$\tgf_{\up\down}(\vr,\vr',\omega_n)$. They satisfy the Dyson
equations (cf. e.g. ref. \cite{MAHAN})
\begin{eqnarray}
\fl\tgg_{\up\up}(\vr,\vr',\omega_n)
&=\tgg^0_{\up\up}(\vr,\vr',\omega_n)\\ \nonumber
 &+ \displaystyle\int d^3s\, d^3s'\; \tgg^0_{\up\up}(\vr,{\bf
 s},\omega_n) [ \Sigma_{\up\up}({\bf s},{\bf
 s}',\omega_n)\tgg_{\up\up}({\bf s}',\vr',\omega_n)\\ \nonumber
 &- W_{\up\down}({\bf s},{\bf s}',\omega_n) \tgf^\dag_{\down\up}({\bf
 s}',\vr',\omega_n)]\quad
  \label{Dyson1}\\
\fl\tgf^\dag_{\down\up}(\vr,\vr',\omega_n) &= \displaystyle\int
 d^3s\, d^3s'\; \tgg^0_{\down\down}({\bf s},\vr,-\omega_n)
 [W_{\down\up}^\dag({\bf s},{\bf s}',\omega_n) \tgg_{\up\up}({\bf
 s}',{\bf s},\omega_n)\\ \nonumber
 &+\Sigma_{\down\down}({\bf s}',{\bf
 s},-\omega_n)\tgf^\dagger_{\down\up}({\bf s},\vr',\omega_n)].
  \label{Dyson2}
\end{eqnarray}
The fermionic Matsubara frequency is as usual
defined by $\omega_n=(2n+1)\pi/\hbar\beta$, where $\beta^{-1}=k_BT$,
with $T$ being the temperature of the system and $k_B$ the Boltzmann
constant. $\Sigma_{\sigma\sigma}$ and $W_{\sigma,-\sigma}$ are the
normal and anomalous self-energies respectively. Analogous equations
to (\ref{Dyson1}) and (\ref{Dyson2}) are satisfied by
$\tgg_{\down\down}$ and $\tgf_{\up\down}$. The assumption of
non-vanishing pair correlations below $T_c$ implies a specific
choice of gauge in order to break the symmetry of the Hamiltonian.
This is implicit in our calculations. We do not need to explicitly
worry about the issue because we are only interested in finding the
critical temperature $T_c$ itself.

In order to further proceed we now need to assume specific
approximations for the self-energies.

\subsection{The normal self-energy}

\label{sec:normal}

In this work we account for the possibility for particles to emit and
reabsorb density and spin fluctuations and adopt the following set of
approximations. We assume $\Sigma_{\up\up}
(\omega_n)=\Sigma^{hf}_{\up\up}+\Sigma^{ph}_{\up\up}(\omega_n)$, and
similarly for $\Sigma_{\up\up}$.  The first term corresponds to the
($\omega$-independent) Hartree self-energy \cite{note1}:
\begin{eqnarray}
\label{Eq:hartree}
\Sigma_{\up\up}^{hf}(\vr,\vr')=-\frac{g}{\beta}\sum_{n}e^{i\omega_n}
\tgg_{\down\down}(\vr,\vr',\omega_n)\delta(\vr-\vr')
\end{eqnarray}
In practice it is generally not needed to consider the full
$\tgg_{\down\down}$ in Eq.~(\ref{Eq:hartree}). It is instead
sufficient to first solve the self-consistent Hartree problem in which
$\tgg_{\down\down}$ is replaced by $\tgg_{\down\down}^{hf}$ (with
$[\tgg_{\down\down}^{hf}]^{-1}=[\tgg_{\down\down}^{0}]^{-1}
-\Sigma_{\down\down}^{hf}$), and then use this set of functions to
calculate the effect of phonons. The self-consistent Hartree problem
is solved by expressing the Hartree Green's functions as
\begin{equation}
\label{Eq:Green} \tgg^{hf}(\vr,\vr',\omega_n)=\sum_{\nu'}
\frac{\phi_{\nu'}(\vr)\phi_{\nu'}^*(\vr')}
{i\hbar\omega_n-\xi_{\nu'}}.
\end{equation}
The quantity $\phi_\nu$ is the wave function of the level $\nu$ in
the Hartree approximation, and $\xi_\nu$ is equal to
$\epsilon_{\nu}^{hf}-\mu$, with $\epsilon_\nu^{hf}$ being the energy
of the level. In Eq. (\ref{Eq:Green}) we omitted the spin indices
since the Hartree Green's function is the same for the two species.
The Hartree self-energy is on the other hand given by
$\Sigma_{\up\up}^{hf}(\vr,\vr')=\delta(\vr-\vr')\sum_\nu
|\phi_\nu(\vr)|^2 n_F(\xi_\nu)$, with $n_F(\xi)=1/(e^{\beta \xi}+1)$
being the Fermi distribution function. The chemical potential is
obtained imposing
\begin{equation}
N=2\sum_\nu \int d^3r\;|\phi_\nu({\bf r})|^2n_F(\xi_\nu).
\end{equation}

\begin{figure}
  \begin{center}
\parbox{10mm}{

\begin{fmfgraph*}(10,20)
  \fmf{fermion, label=$\up$, label.side=right, label.dist=20}{v1,v2}
  \fmf{fermion}{v2,v3}
  \fmf{fermion, label=$\up$, label.side=right, label.dist=20}{v3,v4}
  \fmf{wiggly,right=0.8}{v2,v3}
  \fmfforce{.5w,.05h}{v1}
  \fmfforce{.5w,.3h}{v2}
  \fmfforce{.5w,.7h}{v3}
  \fmfforce{.5w,.95h}{v4}
\end{fmfgraph*}}
\;\;\;=\;\;\;\;\;
\parbox{15mm}{
\begin{fmfgraph*}(15,20)
  \fmf{fermion, label=$\up$, label.side=left, label.dist=20}{v1,v2}
  \fmf{fermion, label=$\up$, label.side=left, label.dist=20}{v2,v3}
  \fmf{fermion, label=$\up$, label.side=left, label.dist=20}{v3,v4}
  \fmf{fermion,left=0.5,label=$\down$, label.side=right,label.dist=15}{w1,w2}
  \fmf{fermion,left=0.5,label=$\down$, label.side=left,label.dist=15 }{w2,w1}
  \fmf{dashes}{v2,w1}
  \fmf{dashes}{v3,w2}
  \fmfforce{.1w,.05h}{v1}
  \fmfforce{.1w,.3h}{v2}
  \fmfforce{.1w,.7h}{v3}
  \fmfforce{.1w,.95h}{v4}
  \fmfforce{.5w,.3h}{w1}
  \fmfforce{.5w,.7h}{w2}
\end{fmfgraph*}}
+\;\;\;\;\;\;
\parbox{20mm}{
\begin{fmfgraph*}(15,35)
  \fmf{fermion, label=$\up$, label.side=left, label.dist=20}{v1,v2}
  \fmf{fermion, label=$\up$, label.side=left, label.dist=20}{v2,v3}
  \fmf{fermion, label=$\up$, label.side=left, label.dist=20}{v3,v4}
  \fmf{fermion, left=0.5,label=$\down$, label.side=right,label.dist=10}{w1,w2}
  \fmf{fermion, left=0.5,label=$\down$, label.side=left,label.dist=10}{w2,w1}
  \fmf{fermion, left=0.5,label=$\up$, label.side=right,label.dist=10}{w3,w4}
  \fmf{fermion, left=0.5,label=$\up$, label.side=left,label.dist=10}{w4,w3}
  \fmf{fermion, left=0.5,label=$\down$, label.side=right,label.dist=10}{w5,w6}
  \fmf{fermion, left=0.5,label=$\down$, label.side=left,label.dist=10}{w6,w5}
  \fmf{dashes}{v2,w1}
  \fmf{dashes}{w2,w3}
  \fmf{dashes}{v3,w6}
  \fmf{dashes}{w4,w5}
  \fmfforce{.1w,.05h}{v1}
  \fmfforce{.1w,.2h}{v2}
  \fmfforce{.1w,.8h}{v3}
  \fmfforce{.1w,.95h}{v4}
  \fmfforce{.5w,.2h}{w1}
  \fmfforce{.5w,.4h}{w2}
  \fmfforce{.8w,.4h}{w3}
  \fmfforce{.8w,.6h}{w4}
  \fmfforce{.5w,.6h}{w5}
  \fmfforce{.5w,.8h}{w6}
\end{fmfgraph*}}
+\;\;\;\;...\;\;\;$(a)$\\
\;\;\;\;\;\;\;\;+\;\;\;\;\;
\parbox{20mm}{
\begin{fmfgraph*}(15,20)
  \fmf{fermion, label=$\up$, label.side=left, label.dist=20}{v1,v2}
  \fmf{fermion, label=$\up$, label.side=left, label.dist=20}{v2,v3}
  \fmf{fermion, label=$\up$, label.side=left, label.dist=20}{v3,v4}
  \fmf{fermion, label=$\down$, label.side=right,label.dist=15}{w2,w1}
  \fmf{fermion,right=0.7, label=$\down$, label.side=right,label.dist=15}{w1,w2}
  \fmf{dashes}{v2,w1}
  \fmf{dashes}{v3,w2}
  \fmfforce{.1w,.05h}{v1}
  \fmfforce{.1w,.3h}{v2}
  \fmfforce{.1w,.7h}{v3}
  \fmfforce{.1w,.95h}{v4}
  \fmfforce{.5w,.3h}{w1}
  \fmfforce{.5w,.7h}{w2}
\end{fmfgraph*}}
+\;\;\;\;\;
\parbox{15mm}{
\begin{fmfgraph*}(15,25)
  \fmf{fermion, label=$\up$, label.side=left, label.dist=20}{v1,v2}
  \fmf{fermion, label=$\up$, label.side=left, label.dist=20}{v2,v3}
  \fmf{fermion, label=$\up$, label.side=left, label.dist=20}{v3,v4}
  \fmf{fermion, label=$\up$, label.side=left, label.dist=20}{v4,v5}
  \fmf{fermion, label=$\down$, label.side=right,label.dist=15}{w2,w1}
  \fmf{fermion, label=$\down$, label.side=right,label.dist=15}{w3,w2}
  \fmf{fermion,right=0.7, label=$\down$, label.side=right,label.dist=15}{w1,w3}
  \fmf{dashes}{v2,w1}
  \fmf{dashes}{v3,w2}
  \fmf{dashes}{v4,w3}
  \fmfforce{.1w,.05h}{v1}
  \fmfforce{.1w,.2h}{v2}
  \fmfforce{.1w,.5h}{v3}
  \fmfforce{.1w,.8h}{v4}
  \fmfforce{.1w,.95h}{v5}
  \fmfforce{.5w,.2h}{w1}
  \fmfforce{.5w,.5h}{w2}
  \fmfforce{.5w,.8h}{w3}
\end{fmfgraph*}}
\;\;\;\;+\;\;\;\;...\;\;\;\;\;$(b)$

 \caption{Phonon-induced self-energy
 $\Sigma^{ph}_{\up\up}(\vr,\vr',\omega_n)$ (Eqs. (\ref{Eq:sigma1}) and
 (\ref{Eq:sigma2})). The continuous lines correspond to Hartree
 Green's functions $\tgg^{hf}(\vr,\vr',\omega_n)$, the dashed lines to
 the bare interaction $g\delta(\vr-\vr')$. The self-energy
 $\Sigma_{\up\up}^{(1)}$ is given by the contributions in $(a)$.
 $\Sigma_{\up\up}^{(2)}$, which arises from coupling to spin phonons,
 is given by the diagrams in $(b)$. The diagrams with an even number
 of bubbles are not included in $(a)$ because the bare interaction
 acts only between atoms in different internal states.}
\label{Fig:Sigma}
  \end{center}
\end{figure}

  We deal with the self-energy $\Sigma_{\up\up}^{ph}$ due to
coupling with collective modes in the context of the Random Phase
Approximation \cite{note2}. We thus suppose $\Sigma_{\up\up}^{ph}$
to be given by the sum of the Feynman graphs shown in Fig.
\ref{Fig:Sigma}, where to each fermion line corresponds a
$\tgg^{hf}$ and to each dashed line an interaction
$g\delta(\vr-\vr')$. This part of self-energy can be written as the
sum of two contributions $\Sigma_{\up\up}^{(1)}$ and
$\Sigma_{\up\up}^{(2)}$ (cf. Fig. \ref{Fig:Sigma} $(a)$ and $(b)$
respectively).  $\Sigma_{\up\up}^{(1)}$ is given by
\begin{eqnarray}
\nonumber
&\Sigma_{\up\up}^{(1)}(\vr,\vr',\omega_n)-\displaystyle\frac{g^2}{\beta}\sum_{\omega_m}
\tgg_{\up\up}^{hf}(\vr,\vr',\omega_n)
\Pi_{\down\down}^{r}(\vr,\vr',\omega_n+\omega_m)
\end{eqnarray}
where $\Pi_{\down\down}^{r}$ is the ring approximation to the
correlation function (for all the correlation functions, and their
relations, used in this Section we refer the reader to
\ref{appcorr}).  This can be written as the sum of a density and a
spin part:
\begin{eqnarray}
\Pi_{\down\down}^{r}(\vr,\vr',\omega_m) =\displaystyle{\frac{1}{4}}
\Big\{\Pi_{\rho}^{r}(\vr,\vr',\omega_m)
+\Pi_{\sigma_z}^{r}(\vr,\vr',\omega_m)\Big\},\;
\end{eqnarray}
with $\omega_m=2m\pi/\hbar\beta$ being a bosonic Matsubara frequency.
The second contribution is instead given by:
\begin{eqnarray}
\nonumber
&\Sigma_{\up\up}^{(2)}(\vr,\vr',\omega_n)=-\displaystyle{\frac{g^2}{4\beta}}
\sum_{\omega_m}\tgg_{\down\down}^{hf}(\vr,\vr',\omega_n)\\
&\Big\{\Pi_{\sigma_x}^{r}(\vr,\vr',\omega_n+\omega_m)
+\Pi_{\sigma_y}^{r}(\vr,\vr',\omega_n+\omega_m)\Big\}.
\end{eqnarray}
Combining the two contributions one finds
\begin{eqnarray}
\nonumber
&\Sigma^{ph}_{\up\up}(\vr,\vr',\omega_n)=-\displaystyle{\frac{g^2}{4\beta}}
\sum_{\omega_m}\tgg^{hf}(\vr,\vr',\omega_n)\\
\nonumber &\times\Big\{\Pi_{\rho}^{r}(\vr,\vr',\omega_n+\omega_m)
+3\Pi_{\sigma_z}^{r}(\vr,\vr',\omega_n+\omega_m)
-2\Pi_{\rho}^{0}(\vr,\vr',\omega_n+\omega_m)\Big\}
\label{Eq:sigma1}\\
\nonumber &= -\displaystyle{\frac{g^2}{\beta}}\sum_{\omega_m}
\tgg^{hf}(\vr,\vr',\omega_n)
\Big\{2\Pi_{\up\up}^{r}(\vr,\vr',\omega_n+\omega_m)\\
&-\Pi_{\up\down}^{r}(\vr,\vr',\omega_n+\omega_m)
-\chi_0(\vr,\vr',\omega_n+\omega_m)\Big\}. \label{Eq:sigma2}
\end{eqnarray}
In Eqs. (\ref{Eq:sigma1}) and (\ref{Eq:sigma2}) we have used
$\Pi_{\sigma_x}=\Pi_{\sigma_y}=\Pi_{\sigma_z}$ and we accounted for
the fact that the single bubble diagram belongs both to
$\Sigma_{\up\up}^{(1)}$ and to $\Sigma_{\up\up}^{(2)}$ and is
therefore doubly counted in a simple summation. For this reason we
subtracted the quantity $2\Pi_{\rho}^{0}$ in Eq. (\ref{Eq:sigma1}).

It is useful to introduce the spectral representations for the correlation
functions:
\begin{equation}
\label{Eq:specpirho} \Pi_{\rho}^{r}(\vr,\vr',\omega_m)
=\int_{0}^{\infty} \frac{d\omega'}{\pi}\;{\rm
Im}[\Pi_{\rho}^{r}(\vr,\vr',\omega')]
\frac{2\omega'}{\omega_m^2+\omega^{'2}},
\end{equation}
\begin{equation}
\label{Eq:specpisigma}
\Pi_{\sigma_z}^{r}(\vr,\vr',\omega_m)=\int_{0}^{\infty}
\frac{d\omega'}{\pi}\;{\rm Im}[\Pi_{\sigma_z}^{r}(\vr,\vr',\omega')]
\frac{2\omega'}{\omega_m^2+\omega^{'2}},
\end{equation}
\begin{equation}
\label{Eq:specpi0} \Pi_{\rho}^0(\vr,\vr',\omega_m)=\int_{0}^{\infty}
\frac{d\omega'}{\pi}\;{\rm Im}[\Pi_{\rho}^{0}(\vr,\vr',\omega')]
\frac{2\omega'}{\omega_m^2+\omega^{'2}},
\end{equation}
where $\Pi_{\rho}^{r}(\vr,\vr',\omega)$,
$\Pi_{\sigma_z}^{r}(\vr,\vr',\omega)$ and
$\Pi_{\rho}^{0}(\vr,\vr',\omega)$ are the retarded correlation
functions (obtained using the analytic continuation $\omega_n\to
\omega+i\eta$), and we have used
$\Pi_{i}^{r}(\vr,\vr',\omega)=-\Pi_{i}^{r}(\vr,\vr',-\omega)$.

Inserting these relations, together with Eq.~(\ref{Eq:Green}), in
Eq.~(\ref{Eq:sigma2}) and carrying out the frequency summation one finds
\begin{eqnarray}
\nonumber &\Sigma_{\up\up}^{ph}(\vr,\vr',\omega_n) =-\displaystyle
g^2\sum_{\nu'}\phi_{\nu'}(\vr)\phi_{\nu'}^*(\vr')
\int_0^{\infty} \frac{d\omega'}{\pi}\;\\
\nonumber &{\rm Im}\big[2\Pi_{\up\up}^{r}(\vr,\vr',\omega')-
\Pi_{\up\down}^{r}(\vr,\vr',\omega')
-\chi_0(\vr,\vr',\omega')\big]\\
&\times\left(\displaystyle{\frac{n_B(\omega')+1-n_F(\xi_{\nu'})}
{i\omega_n-\hbar^{-1}\xi_{\nu'}-\omega'}+
\frac{n_B(\omega')+n_F(\xi_{\nu'})}
{i\omega_n-\hbar^{-1}\xi_{\nu'}+\omega'}}\right),\quad
\end{eqnarray}
where $n_B(\omega)=1/(e^{\beta \hbar\omega}-1)$ is the Bose distribution
function.

The self-energy of the quasi-particle in the level $\nu$ is given by
$\Sigma^{ph}_\nu(\omega_n)=\int\,d^3r\,d^3r'\,\phi^*_\nu(\vr)\phi_\nu(\vr')
\Sigma^{ph}(\vr,\vr',\omega_n)$ and therefore
\begin{eqnarray}
\nonumber & \Sigma^{ph}_\nu(\omega_n)= -\displaystyle{\sum_{\nu'}
  \int_0^\infty \frac{d\omega'}{\pi} \;\sigma_{\nu\nu'}(\omega') }\\
&\times {\left(\displaystyle{\frac{1+n_B(\omega')-n_F(\xi_{\nu'})}
    {i\omega_n-\hbar^{-1}\xi_{\nu'}-\omega'}
    +\frac{n_B(\omega')+n_F(\xi_{\nu'})}
    {i\omega_n-\hbar^{-1}\xi_{\nu'}+\omega'}}\right)},
    \label{sigma}
\end{eqnarray}
where
\begin{eqnarray}
\nonumber \sigma_{\nu\nu'}(\omega)&=g^2\displaystyle\int d^3r
d^3r'\; \phi_{\nu'}(\vr)\phi_\nu^*(\vr)\phi_{\nu'}^*(\vr')
\phi_\nu(\vr')\\
&\times{\rm Im}\big[2\Pi_{\up\up}^{r}(\vr,\vr',\omega)-
\Pi_{\up\down}^{r}(\vr,\vr',\omega)
-\chi_0(\vr,\vr',\omega)\big].\quad
\end{eqnarray}

\subsection{The anomalous self-energy}

\label{sec:anomalous}

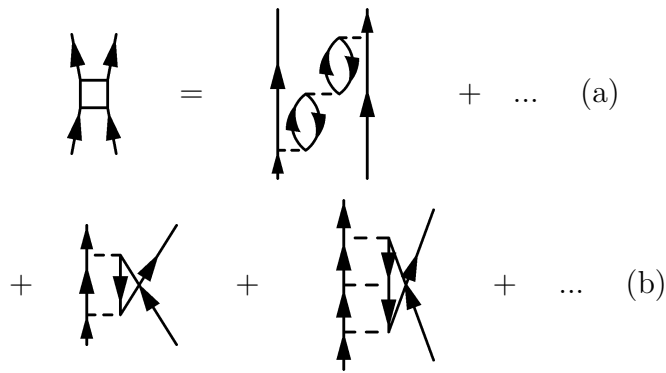
\begin{figure}
  \begin{center}
\parbox{15mm}{
  \begin{fmfgraph*}(10,20)
    \fmfforce{.2w,.1h}{v1}
    \fmfforce{.8w,.1h}{v2}
    \fmfforce{.2w,.9h}{v3}
    \fmfforce{.8w,.9h}{v4}
    \fmfpolyn{empty,tension=1.}{G}{4}
    \fmf{fermion, label=$\up$, label.side=left, label.dist=20}{v1,G1}
    \fmf{fermion, label=$\down$, label.side=right, label.dist=20}{v2,G2}
    \fmf{fermion}{G4,v3}
    \fmf{fermion}{G3,v4}
\end{fmfgraph*}}
=\;\;\;\;\;\;
\parbox{20mm}{
\begin{fmfgraph*}(15,25)
  \fmf{fermion, label=$\up$, label.side=left, label.dist=20}{v1,v2}
  \fmf{fermion, label=$\up$, label.side=left, label.dist=20}{v2,v3}
  \fmf{fermion, label=$\down$, label.side=right, label.dist=20}{v4,v5}
  \fmf{fermion, label=$\down$, label.side=right, label.dist=20}{v5,v6}
  \fmf{fermion,left=0.5, label=$\down$, label.side=right, label.dist=10}{w1,w2}
  \fmf{fermion,left=0.5, label=$\down$, label.side=left, label.dist=10}{w2,w1}
  \fmf{fermion,left=0.5, label=$\down$, label.side=left, label.dist=10}{w3,w4}
  \fmf{fermion,left=0.5, label=$\down$, label.side=right, label.dist=10}{w4,w3}
  \fmf{dashes}{v2,w1}
  \fmf{dashes}{w2,w3}
  \fmf{dashes}{w4,v5}
  \fmfforce{.1w,.05h}{v1}
  \fmfforce{.1w,.2h}{v2}
  \fmfforce{.1w,.95h}{v3}
  \fmfforce{.9w,.05h}{v4}
  \fmfforce{.9w,.8h}{v5}
  \fmfforce{.9w,.95h}{v6}
  \fmfforce{.35w,.2h}{w1}
  \fmfforce{.35w,.5h}{w2}
  \fmfforce{.65w,.5h}{w3}
  \fmfforce{.65w,.8h}{w4}

\end{fmfgraph*}}
\;\;\;+\;\;\;...\;\;\;\;(a)\\

+\;\;\;\;
\parbox{20mm}{
\begin{fmfgraph*}(15,20)
  \fmf{fermion, label=$\up$, label.side=left, label.dist=20}{v1,v2}
  \fmf{fermion, label=$\up$, label.side=left, label.dist=20}{v2,v3}
  \fmf{fermion, label=$\up$, label.side=left, label.dist=20}{v3,v4}
  \fmf{fermion, label=$\down$, label.side=right, label.dist=20}{v5,w2}
  \fmf{fermion}{w1,v6}
  \fmf{fermion, label=$\down$, label.side=right, label.dist=15}{w2,w1}
  \fmf{dashes}{v2,w1}
  \fmf{dashes}{w2,v3}
  \fmfforce{.1w,.1h}{v1}
  \fmfforce{.1w,.3h}{v2}
  \fmfforce{.1w,.7h}{v3}
  \fmfforce{.1w,.9h}{v4}
  \fmfforce{.9w,.1h}{v5}
  \fmfforce{.9w,.9h}{v6}
  \fmfforce{.4w,.3h}{w1}
  \fmfforce{.4w,.7h}{w2}
\end{fmfgraph*}}
+\;\;\;\;\;\;\;
\parbox{20mm}{
\begin{fmfgraph*}(15,25)
  \fmf{fermion, label=$\up$, label.side=left, label.dist=20}{v1,v2}
  \fmf{fermion, label=$\up$, label.side=left, label.dist=20}{v2,v7}
  \fmf{fermion, label=$\up$, label.side=left, label.dist=20}{v7,v3}
  \fmf{fermion, label=$\up$, label.side=left, label.dist=20}{v3,v4}
  \fmf{fermion, label=$\down$, label.side=right, label.dist=20}{v5,w2}
  \fmf{fermion}{w1,v6}
  \fmf{fermion, label=$\down$, label.side=right, label.dist=15}{w2,w3}
  \fmf{fermion, label=$\down$, label.side=right, label.dist=15}{w3,w1}
  \fmf{dashes}{v2,w1}
  \fmf{dashes}{w2,v3}
  \fmf{dashes}{v7,w3}
  \fmfforce{.1w,.05h}{v1}
  \fmfforce{.1w,.25h}{v2}
  \fmfforce{.1w,.75h}{v3}
  \fmfforce{.1w,.95h}{v4}
  \fmfforce{.9w,.1h}{v5}
  \fmfforce{.9w,.9h}{v6}
  \fmfforce{.1w,.5h}{v7}
  \fmfforce{.5w,.25h}{w1}
  \fmfforce{.5w,.75h}{w2}
  \fmfforce{.5w,.5h}{w3}
\end{fmfgraph*}}
+\;\;\;\;...\;\;\;\;(b)

\caption{Phonon-induced interaction
  $V^{ind}_{\up\down}(\vr,\vr',\omega_n)$ (see Eq. (\ref{vind})). The
  diagrams with an odd numbers of bubbles are not included in $(a)$
  because we are considering only the interaction between
  quasi-particles with opposite spin states.}
\label{Fig:Vind}
  \end{center}
\end{figure}

The anomalous self-energy  in
Eqs. (\ref{Dyson1}) and (\ref{Dyson2}) is given by (cf. \cite{MAHAN})
\begin{eqnarray}
  \nonumber
 & W_{\up\down}(\vr,\vr',\omega_n)=-\displaystyle\frac{1}{\beta}\sum_{\omega_m}
 V^{eff}_{\up\down}(\vr,\vr',\omega_m)\tgf_{\up\down}(\vr,\vr',\omega_n+\omega_m),
  \label{Eq:anomalous}
\end{eqnarray}
where $V^{eff}_{\up\down}$ is an effective interaction between
quasi-particles in states $\up$ and $\down$.  Accounting also for
the possibility for particles to interact exchanging density and
spin fluctuations, $V_{\up\down}^{eff}(\vr,\vr',\omega_m)$ is the
sum of two terms: the ($\omega$-independent) bare interaction
$g\delta(\vr-\vr')$ and the induced interaction
\begin{eqnarray}
\label{vind}
V^{ind}_{\up\down}(\vr,\vr',\omega_m)=\displaystyle{\frac{g^2}{4}}
\big[\Pi_\rho^{r}(\vr,\vr',\omega_m)
-3\Pi_{\sigma_z}^{r}(\vr,\vr',\omega_m)\big]
\end{eqnarray}
which corresponds to the diagrams shown in Figure \ref{Fig:Vind}.

$\Pi_\rho^{r}$ is related to the exchange of density modes,
$\Pi_{\sigma_i}^{r}$ to the exchange of spin modes. Introducing Eqs.
(\ref{Eq:specpirho}) and (\ref{Eq:specpisigma}) in
Eq.~(\ref{vind}) yields
\begin{eqnarray}
\nonumber
V^{ind}_{\up\down}(\vr,\vr',\omega_m)=\frac{g^2}{4}\int_{0}^{\infty}
\frac{d\omega'}{\pi}
\frac{2\omega'}{\omega_m^2+{\omega'}^{2}}\\
\times {\rm Im}\big[\Pi_\rho^{r}(\vr,\vr',\omega')
-3\Pi_{\sigma_z}^{r}(\vr,\vr',\omega')\big]. \label{specV}
\end{eqnarray}
Since the imaginary parts of the retarded correlation functions are
always negative, we see that the density modes induce an effective
attractive interaction, while the spin modes induce a repulsive one.
In the Hartree-Fock basis the anomalous self-energy, as well as
$\tgf$, depends in principle on two indices $\nu_1$ and $\nu_2$ as
$W_{\nu_1\nu_2}(\omega_n)=\int d^3r d^3r'
\;\phi_{\nu_1}(\vr)\phi_{\nu_2}(\vr')W_{\up\down}(\vr,\vr',\omega_n)$.
In the same basis, Eq.~(\ref{Eq:anomalous}) becomes:
\begin{eqnarray}
  \nonumber & W_{\nu_1\nu_2}(\omega_n)=-\displaystyle\frac{1}{\beta}
 \sum_{\nu_3\nu_4,\omega_m} V^{eff}_{\nu_1\nu_2,\nu_3\nu_4}(\omega_m)
 \tgf_{\nu_3\nu_4}(\omega_n+\omega_m).
  \label{Eq:anomalous2}
\end{eqnarray}
where
\begin{eqnarray}
\nonumber V^{eff}_{\nu_1\nu_2,\nu_3\nu_4}(\omega_m)=\int d^3rd^3r'\;
 \phi_{\nu_1}(\vr)\phi_{\nu_2}(\vr')
 \phi_{\nu_3}^*(\vr)\phi_{\nu_4}^*(\vr')
 V^{eff}_{\up\down}(\vr,\vr',\omega_m)
  \label{Eq:elmatint}
\end{eqnarray}
Solving the Dyson equations with the full $\nu$ dependence is rather
complicated.  A consistent simplification may be attained by keeping
only the matrix elements between states connected by the time
reversal ($\phi_{\nu_2}=\phi_{\nu_1}^*$ and
$\phi_{\nu_4}=\phi_{\nu_3}^*$). This is in keeping with the fact
that those matrix elements are the largest because the overlap of
the wavefunctions in Eq.~(\ref{Eq:elmatint}) is maximal (see for
instance Ref. \cite{BRUUN1,BRUUN2}). With this assumption both the
anomalous self-energy and the anomalous Green's function depend on a
single index and Eq.~(\ref{Eq:anomalous2}) becomes:
\begin{equation}
  \label{Eq:delta}
 W_\nu(\omega_n)=-\frac{1}{\beta}\sum_{\nu'\omega_{m}}
V_{\nu\nu'}^{eff}(\omega_{m})\tgf_{\nu'}(\omega_{n}+\omega_m),
\end{equation}
with
\begin{equation}
\label{Eq:int}
V_{\nu\nu'}^{eff}(\omega_m)=g_{\nu\nu'}+
\displaystyle \int_{0}^{\infty}\frac{d\omega'}{\pi}\;
v_{\nu\nu'}(\omega')\frac{2\omega'}{\omega_m^2+{\omega'}^{2}},
\end{equation}
$g_{\nu\nu'}$ being the matrix element of the bare interaction and
\begin{eqnarray}
\fl \nonumber v_{\nu\nu'}(\omega)=\displaystyle\frac{g^2}{4}\int
d^3r d^3r'
\phi_{\nu}(\vr)\phi_{\nu}^*(\vr')\phi^*_{\nu'}(\vr)\phi_{\nu'}(\vr')
{\rm Im}\bigl[\Pi_\rho^{r}(\vr,\vr',\omega)
-3\Pi_{\sigma_z}^{r}(\vr,\vr',\omega)\bigr]. \label{Eq:matelv}
\end{eqnarray}

The Dyson equations Eq. (\ref{Dyson1}) and
(\ref{Dyson2}) on the other hand take the simple form:
\begin{eqnarray}
  \tgg_\nu(\omega_n)
  =G_\nu(\omega_n)-G_\nu(\omega_n)W_\nu(\omega_n)
\tgf_\nu^\dag(\omega_n),\label{Eq:dyson41}
\end{eqnarray}
and
\begin{eqnarray}
  \tgf^\dag_\nu(\omega_n)
  =G_\nu(-\omega_n)W_\nu^\dagger(\omega_n)\tgg_\nu(\omega_n),
\label{Eq:dyson42}
\end{eqnarray}
where
\begin{equation}
  \label{Eq:G}
  G_{\nu}(\omega_{n})  \frac{1}{i\omega_n-\xi_\nu-\Sigma^{ph}_\nu(\omega_n)}.
\end{equation}

Close to $T_c$ one can keep only the linear contribution of the
anomalous self-energy and obtain for the anomalous propagator
\begin{equation}
\label{Eq:F}
\tgf_{\nu}(\omega_{n})W_{\nu}(\omega_n)G_{\nu}(\omega_n)G_{\nu}(-\omega_n).
\end{equation}

\subsection{One pole approximation and quasi-particle properties}

\label{sec:onepole}

We now proceed by expanding the propagators close to the poles which
correspond to the quasi-particle excitation energies.  As will be shown below,
this approximation holds in the weak and likely also in the medium coupling
regimes. By weak coupling we mean $\lambda \lesssim 0.2$ and by medium
coupling $0.2\lesssim \lambda \lesssim 0.5$. In order to carry out the one
pole approximation on the finite temperature propagators (the $T=0$ case is
treated e. g. in Ref. \cite{BALDO}), one needs to
introduce the spectral representation
\begin{equation}
    \label{lehmann}
    G_\nu(\omega_n)=\int_{-\infty}^\infty
    \frac{d\omega'}{2\pi}\,\frac{\rho_\nu(\omega')}{i\omega_n-\omega'}
\end{equation}
where the spectral density $\rho$ is given by
\begin{eqnarray}
    \label{rho}
    \rho_\nu(\omega)    -i\bigl[ G_\nu^{A}(\omega)-G_\nu^{R}(\omega)\bigr]
\end{eqnarray}
with
$G_\nu^{A}(\omega)= G_\nu(\omega_n)|_{\omega_n\rightarrow\omega-i\eta}$ and
$G_\nu^{R}(\omega)= G_\nu(\omega_n)|_{\omega_n\rightarrow\omega+i\eta}$
being the advanced and retarded correlation functions respectively.
The corresponding self-energies satisfy the relations
\begin{eqnarray}
  {\rm Im}\Sigma^{ph,\,A}_\nu(\omega)=-{\rm
    Im}\Sigma^{ph,\,R}_\nu(\omega)&>&0 \;\; \forall\omega
\end{eqnarray}
and
\begin{eqnarray}
  {\rm Re}\Sigma^{ph,\, A}_\nu(\omega)={\rm Re}\Sigma^{ph,\, R}_\nu
  (\omega)\;\; \forall\omega.
\end{eqnarray}
If $|{\rm Im}\Sigma^{ph}_\nu(\omega)|\ll |\xi_\nu+ {\rm
Re}\Sigma^{ph}_\nu(\omega)|$, the inverse advanced Green's function is
approximately given by
\begin{eqnarray}
  [G_{\nu}^A(\omega)]^{-1}
  \simeq Z_\nu^{-1}(\omega-\epsilon_\nu)
  -i\,{\rm Im}\Sigma^{ph,\,A}_\nu(\epsilon_\nu),
\end{eqnarray}
where $\epsilon_\nu$ satisfies the relation $\epsilon_\nu=\xi_\nu+{\rm Re}
\Sigma_\nu^{ph}(\epsilon_\nu)$. The analogous expression holds for the
retarded Green's function.  With the definition $\gamma_\nu={\rm
Im}\Sigma^{ph,\,A}_\nu(\epsilon_\nu)$ the spectral function becomes
\begin{equation}
  \rho_\nu(\omega)=2\,Z_\nu\frac{\gamma_\nu}{(\omega-\epsilon_\nu)^2
    +\gamma_\nu^2},
\end{equation}
with the characteristic lorentzian shape of width $\gamma_\nu$
centered about the renormalized quasi-particle energy $\epsilon_\nu$.
The factor $Z_\nu$ is defined by
\begin{equation}
Z_\nu=\left(1-\frac{\partial{\rm Re}\Sigma^{ph}_\nu(\omega)}{\partial
(\hbar\omega)}\Big|_{\hbar\omega=\epsilon_\nu}\right)^{-1}\leq1,
\end{equation}
and it physically represents the strength of the lorentzian
quasi-particle peak, as it is shown by the relation:
\begin{equation}
    Z_\nu=\int_{-\infty}^\infty \frac{d\omega'}{2\pi}\,
    \rho_\nu(\omega').
\end{equation}
Substitution into Eq. (\ref{lehmann})  finally gives
\begin{eqnarray}
  G_\nu(\omega_n)\simeq
  Z_\nu\left[\displaystyle\frac{\theta(\omega_n)}
  {i\hbar\omega_n-\epsilon_\nu+i\gamma_\nu}+
  \displaystyle\frac{\theta(-\omega_n)}
  {i\hbar\omega_n-\epsilon_\nu-i\gamma_\nu}\right]\quad
      \label{Gren}
\end{eqnarray}
(see Ref. \cite{MOREL}). The possibility of performing the one pole
approximation relies on the assumption that the imaginary part of the
phonon-induced self-energy is small as compared to the quasi-particle
energy in the proximity of the Fermi surface.  This is equivalent to
requiring the quasi-particle lifetime to be long.  The discussion on
the validity of this assumption will be presented in Sect. III, where
we will show the results for the self-energy calculated using
Eq.~(\ref{sigma}).  These are also important in order to determine in
which range of the interaction parameter the description of Fermi
gases in terms of quasi-particles is well suited.

\subsection{The gap equation}

\label{sec:gap}

Using Eq.~(\ref{Gren}) the anomalous propagator in Eq.~(\ref{Eq:F}) becomes:
\begin{eqnarray}
    \nonumber
    \tgf_\nu(\omega_n)\simeq\displaystyle\frac{Z_\nu^2
    \,W_\nu(\omega_n)}{2\epsilon_\nu}\Bigl[\displaystyle
    \frac{1}{i|\hbar\omega_n|-\epsilon_\nu+i\gamma_\nu} -\displaystyle
    \frac{1}{i|\hbar\omega_n|+\epsilon_\nu+i\gamma_\nu} \Bigr].
  \label{Fren}
\end{eqnarray}
It allows a spectral representation itself:
\begin{equation}
    \label{lehmannF}
    \tgf_\nu(\omega_n)=\int_{-\infty}^\infty
    \frac{d\omega'}{2\pi}\,\frac{f_\nu(\omega')}{i\omega_n-\omega'},
\end{equation}
with
\begin{eqnarray}
  \nonumber
     f_\nu(\omega)=\displaystyle{
    \frac{Z_\nu^2 \,W_\nu(\omega)}{2\epsilon_\nu}
    \Bigl[ \frac{2\gamma_\nu}{(\omega-\epsilon_\nu)^2+\gamma_\nu^2}}
   -\displaystyle{\frac{2\gamma_\nu}{(\omega+\epsilon_\nu)^2
    +\gamma_\nu^2}}\Bigr].
  \label{eq:specF}
\end{eqnarray}
Using the one pole approximation for Eq. (\ref{eq:specF}) and the
symmetry of the gap function $W_\nu(\omega)=W_\nu(-\omega)$
Eq.~(\ref{Fren}) becomes
\begin{eqnarray}
  \nonumber
    \tgf_\nu(\omega_n)    \displaystyle{\frac{Z_\nu\,\Delta_\nu}{2\epsilon_\nu}
    \Bigl[ \frac{1}{i|\hbar\omega_n|-\epsilon_\nu+i\gamma_\nu}}
    -\frac{1}{i|\hbar\omega_n|+\epsilon_\nu+i\gamma_\nu} \Bigr].
    \label{Eq:frenfin}
\end{eqnarray}
where $\Delta_\nu=Z_\nu W_\nu(\epsilon_\nu)$.  The gap equation now
follows by inserting Eqs. (\ref{Eq:int}) and (\ref{Eq:frenfin}) into
Eq.~(\ref{Eq:delta})
\begin{eqnarray}
  \nonumber
  & W_\nu(\omega_n)=-\displaystyle\frac{1}{\beta}\sum_{\nu'\omega_{n'}}
  \frac{Z_{\nu'}\,\Delta_{\nu'}}{2\epsilon_{\nu'}}
  \label{Eq:gapgamma}
  \left\{
  g_{\nu\nu'}+\displaystyle\int_{0}^{\infty}\frac{d\omega}{\pi}\;
  v_{\nu\nu'}(\omega)
  \frac{2\omega}{(\omega_n-\omega_{n'})^2+\omega^{2}}\right\}\\
  &\nonumber \displaystyle{ \times\left\{
  \frac{1}{i|\hbar\omega_{n'}|-\epsilon_{\nu'}+i\gamma_{\nu'}}-
  \frac{1}{i|\hbar\omega_{n'}|+\epsilon_{\nu'}+i\gamma_{\nu'}}
  \right\}}.
\end{eqnarray}
with $\omega_{n'}=\omega_n+\omega_m$.  Neglecting the quasi-particle
width and carrying out the Matsubara sum one gets
\begin{eqnarray}
  \nonumber & W_\nu(\omega_n)\simeq \displaystyle\sum_{\nu'}
  \frac{Z_{\nu'}\,\Delta_{\nu'}}{2\epsilon_{\nu'}}(1-2n_F(\epsilon_{\nu'}))\\
  \nonumber &\times\left\{ g_{\nu\nu'}
  +\displaystyle\int_{0}^{\infty}\frac{d\omega}{\pi}
  v_{\nu\nu'}(\omega)
  \displaystyle{\left[\frac{1}{i\omega_n+|\epsilon_{\nu'}|+\omega}
  +\frac{1}{-i\omega_n+|\epsilon_{\nu'}|+\omega}\right]}\right\}.
\end{eqnarray}
After the analytic continuation $i\omega_n\to \epsilon_{\nu}$ and
multiplication by $Z_{\nu}$ one finds:
\begin{eqnarray}
  \nonumber & \Delta_\nu\simeq \displaystyle\sum_{\nu'}
  \frac{\Delta_{\nu'}}{2\epsilon_{\nu'}}(1-2n_F(\epsilon_{\nu'}))\\
  \nonumber &\times\left\{\bar g_{\nu\nu'}
  +\displaystyle\int_{0}^{\infty}\frac{d\omega}{\pi} \bar
  v_{\nu\nu'}(\omega)
  \displaystyle{\left[\frac{1}{\epsilon_{\nu}+|\epsilon_{\nu'}|+\omega}
  +\frac{1}{-\epsilon_{\nu}+|\epsilon_{\nu'}|+\omega}\right]}\right\}.
  \label{Eq:gapfinalnogamma}
\end{eqnarray}
with $\bar g_{\nu\nu'}=Z_\nu g_{\nu\nu'}Z_{\nu'}$ and similarly for
$\bar v_{\nu\nu'}$. The terms in the curly brackets constitute an
effective interaction matrix element between quasi-particles in states
$\nu$ and $\nu'$. The second term in particular has the form of an
induced interaction obtained from second order perturbation theory
with a fermion-boson Hamiltonian of the form:
\begin{eqnarray}
\nonumber
&\hat H = \displaystyle\sum_{\nu\sigma} \epsilon_{\nu}\;
\hat a_{\nu\sigma}^\dagger
\hat a_{\nu\sigma}+
\displaystyle\sum_{\omega s} \hbar\omega\; \hat b_{\omega s}^\dagger
\hat b_{\omega s}\\
\nonumber
&+\displaystyle\frac{1}{2}\sum_{\nu\nu'\sigma}
\bar g_{\nu\nu'} a_{\nu\sigma}^\dagger
a_{\bar\nu'\bar\sigma}^\dagger a_{\bar\nu\bar\sigma} a_{\nu\sigma}\\
&+\displaystyle\sum_{\nu\sigma,\nu'\sigma';\omega s}
\sqrt{\bar v_{\nu\nu'}(\omega)}\;\;
\hat a_{\nu\sigma}^\dagger \hat b_{\omega s}^\dagger
\hat a_{\nu'\sigma'}+ {\rm c.c.}
\label{Eq:B-FHam}
\end{eqnarray}
$\sqrt{\bar v_{\nu\nu'}(\omega)}$ being the effective
particle-vibration coupling \cite{BOHRMOTTELSON}. In the equation
$\bar\sigma=\up$ if $\sigma=\down$ and viceversa, and $s$ is the index
associated with the spin carried by the phonon.

To perform the numerical calculations one may replace the incoming
energies $\epsilon_\nu$ at the denominators of
Eq.~(\ref{Eq:gapfinalnogamma}) with their moduli.  The corresponding
equation is only approximately correct and coincides with the one
obtained within the Bloch-Horowitz perturbation theory using the
Hamitonian in Eq.~(\ref{Eq:B-FHam}) \cite{GORI}.  With this
replacement though, the denominators in
Eq.~(\ref{Eq:gapfinalnogamma}) never vanish and the framework is
therefore well suited to obtain numerical results. The two
formulas on the other hand necessarily lead to very close
predictions for $T_c$ since they coincide at the Fermi surface,
$\epsilon_\nu=0$, where pairing is most important.

In the case of a frequency independent interaction
Eq.~(\ref{Eq:gapgamma}) has a simple solution also including the level
width. This case was treated in Ref. \cite{MOREL}.
For instance neglecting the
phonon-induced interaction in Eq.~(\ref{Eq:gapgamma}) one obtains
\begin{eqnarray}
\nonumber & \Delta_\nu= \displaystyle\sum_{\nu'}
\bar g_{\nu\nu'}\frac{\Delta_{\nu'}}{2\epsilon_{\nu'}}
h(\epsilon_{\nu'},\gamma_{\nu'},T),
\label{Eq:gapsigamma}
\end{eqnarray}
where
\begin{eqnarray}
h(\epsilon_\nu,\gamma_\nu,T)=\frac{2}{\beta}
\displaystyle\sum_{n=0}^\infty
\left[\frac{2\epsilon_\nu}{(\hbar\omega_n+\gamma_\nu)
+\epsilon_\nu^2}\right].
\end{eqnarray}
As a first refinement of the theory one can combine Eq.~(\ref{Eq:gapsigamma})
and the Bloch-Horowitz version of Eq.~(\ref{Eq:gapfinalnogamma}) to get
\begin{eqnarray}
\fl \nonumber \Delta_\nu= \displaystyle\sum_{\nu'}
\frac{\Delta_{\nu'}}{2\epsilon_{\nu'}} \Bigg\{\bar g_{\nu\nu'}\;
h(\epsilon_{\nu'},\gamma_{\nu'},T)
\displaystyle\left[\int_{0}^{\infty}\frac{d\omega}{\pi}\;
\displaystyle\frac{\bar v_{\nu\nu'}(\omega)}
{|\epsilon_{\nu}|+|\epsilon_{\nu'}|+\omega}\right]
[1-2n_F(\epsilon_{\nu'})]\Bigg\}. \label{Eq:gapfinalsigamma}
\end{eqnarray}
For a spherically symmetric system the gap, as well as all the single
particle properties, are independent of the magnetic quantum number
$m$ and the equation can be further simplified.

\subsection{Multipolar expansion}

\label{sec:multi}

Due to the spherical symmetry, the Hartree wavefunctions take the
form $\phi_\nu(\vr)= R_{nl}(r)Y_{lm}(\Omega)$. Substituting in
addition the multipolar expansion for the response function (see
Eq.~(\ref{multipol})) the angular integrals in Eq.~(\ref{Eq:matelv})
become
\begin{eqnarray}
\nonumber
&\int d\Omega\; Y_{lm}(\Omega)Y_{l'm'}^*(\Omega)Y_{LM}(\Omega)=\\
&\times (-1)^{l'-m'}\langle lm||Y_{LM} ||l'm'\rangle
\left(\begin{array}{ccc}
l & l' & L\\
m & m' & M
\end{array}
\right)
\end{eqnarray}
and the same result holds for its complex conjugate. Here
\begin{eqnarray}
\nonumber \left<lm||Y_{LM} ||l'm'\right>&=&
\left[\frac{(2l+1)(2L+1)(2l'+1)}{4\pi}\right]^{1/2}
\left(\begin{array}{ccc}
l & l' & L\\
0 & 0 & 0
\end{array}
\right)
\end{eqnarray}
is the reduced matrix element and
\begin{equation}
\left(\begin{array}{ccc}
l & l' & L\\
m & m' & M
\end{array}
\right)
\end{equation}
is the Wigner 3-$j$ symbol.

Recalling that
\begin{equation}
\displaystyle\sum_{mm'M}
\left(\begin{array}{ccc}
l & l' & L\\
m & m' & M
\end{array}
\right)^2=1
\end{equation}
the gap equation Eq.~(\ref{Eq:gapfinalsigamma}) can finally be
written in the rotationally invariant form
\begin{eqnarray}
  \nonumber & \Delta_{nl}= \displaystyle\sum_{n'l'}
  \frac{\Delta_{n'l'}}{2\epsilon_{n'l'}}
  \sqrt{\frac{2l'+1}{2l+1}}\bigg\{\bar
  g_{nl,n'l'}^{(00)}h(\epsilon_{n'l'},\gamma_{n'l'},T) \\ &
  +\left[\displaystyle\frac{1}{\pi}\int_{0}^{\infty}d\omega\;
  \displaystyle\frac{\bar v_{nl,n'l'}^{(00)}(\omega)}
  {|\epsilon_{nl}|+|\epsilon_{n'l'}|+\omega}\right][1-2n_F(\epsilon_{n'l'})]
  \bigg\}.\quad
\label{Eq:gapfinalsigamma2}
\end{eqnarray}
$\bar v_{nl,n'l'}^{(00)}(\omega)$ is
the matrix element coupled to zero angular momentum \cite{BOHRMOTTELSON}
and is given by
\begin{eqnarray}
\nonumber & v_{nl,n'l'}^{(00)}(\omega)=\displaystyle\frac{g^2}
{\sqrt{2l+1}\sqrt{2l'+1}}\sum_{L=|l'-l|}^{l'+l}
\displaystyle\int dr\,r^2\; dr'\;{r'}^2 f_{ll'L}(r)f_{ll'L}(r')\\
&\times {\rm Im}\{\Pi_{L\;\uparrow\uparrow}(r,r',\omega)
-2\Pi_{L\;\uparrow\downarrow}(r,r,\omega)\},
\end{eqnarray}
with $f_{ll'L}(r)=\langle lm||Y_{LM} ||l'm'\rangle R_{n'l'}(r)R_{nl}(r)$.
An analogous definition holds for $\bar g_{nl,n'l'}^{(00)}$.

The same multipolar expansion can also be applied to the calculation
of the self-energy. Eq.~(\ref{sigma}) then becomes:
\begin{eqnarray}
\nonumber
&  \Sigma^{ph}_{nl}(\omega_n)= \sum_{n'l'}
\displaystyle{
\sqrt{\frac{2l'+1}{2l+1}}
\int_0^\infty\frac{dE}{2\pi}} \;\sigma_{nl,n'l'}^{(00)}(E) \\
&\times  {\left(\displaystyle{\frac{1+n_B(E)-n_F(\xi_{n'l'})}
  {i\omega_n-\hbar^{-1}\xi_{n'l'}-E}
  +\frac{n_B(E)+n_F(\xi_{n'l'})}
  {i\omega_n-\hbar^{-1}\xi_{n'l'}+E}}\right)}\nonumber \\
    \label{sigma2}
\end{eqnarray}
where here
\begin{eqnarray}
\nonumber &\sigma_{nl,n'l'}^{(00)}= \displaystyle\frac{g^2}
{\sqrt{2l+1}\sqrt{2l'+1}}
\sum_{L=|l'-l|}^{l'+l}
\displaystyle\int dr\,r^2\; dr'\;{r'}^2 f_{ll'L}(r)f_{ll'L}(r')\\
&\times {\rm Im}\{2\Pi_{L\;\uparrow\uparrow}(r,r',\omega)
-\Pi_{L\;\uparrow\downarrow}(r,r,\omega)
-\chi_{L\;\uparrow\downarrow}(r,r,\omega)\}.\nonumber \\
\end{eqnarray}

Eq.~(\ref{Eq:gapfinalsigamma2}) is our reference equation for the
evaluation of $T_c$. It is an eigenvalue equation for the
eigenvector $\Delta_{nl}$. The critical temperature $T_c$ is the
highest temperature at which the operator on the RHS has eigenvalue
1. The equation contains a sum over $n'$ which needs special care.
The contribution of the bare interaction, in fact, leads to an
ultra-violet divergence, since it arises from a contact
approximation to the true interatomic potential. Several approaches
have been developed to regularize this divergence. They involve a
renormalization of the coupling constant $g$, which is eliminated in
favor of the low energy $t$-matrix ($t=4\pi\hbar^2a_F/m$) by means
of the Lippmann-Schwinger equation, or the introduction of a contact
pseudopotential $g\delta({\bf r})[(\partial/\partial r) r]\cdot$
\cite{RENORM1,RENORM2}.  For a trapped gas, for which only the bare
interaction is considered, the results of these calculations may be
reproduced by directly replacing $g$ with the $t$-matrix, provided
one simultaneously introduces a cut-off in the sum at
$\epsilon_{n'l'}=\epsilon_F$ \cite{BRUUN1,BRUUN2}. Because the main
scope of the present work is on the phonon-induced effects, we treat
the bare interaction at this level of approximation.  Due to the
explicit dependence on $\epsilon_{n'l'}$ instead, the phonon-induced
interaction has a natural cut-off. Numerical calculations have shown
that the terms with $|\epsilon_{n'l'}|\gtrsim\epsilon_F$ give
negligible contributions.

\section{Results}

\label{sec:results}

In this Section we present the results of our numerical calculations
on the quasi-particle properties and the critical temperature. As we
mentioned in the introduction, the calculations were carried out for
two specific cases: a spherically symmetric infinite square well, and
a 3D isotropic harmonic oscillator. The infinite square well case is
interesting for two reasons. First of all, it has very
similar properties to those of an infinite homogeneous system, and the
results of numerical calculations can therefore be compared with the
predictions of the analytical or semi-analytical formulas. Second, a
detailed comparison with the harmonically trapped case allows to extract
the important physical consequences of the discrete shell structure
induced by the harmonic trapping. The results are shown first for the
spherical well, and then for the harmonically confined system.  In
both cases the calculations were carried out with a number of particles
fixed to $N=1000$ and three different values of the interaction strength
$\lambda$ defined in the introduction.

\subsection{Infinite square well and uniform system}

\label{sec:resunif}

We here consider a system of 1000 particles in a spherically symmetric
infinite square well of radius $R$. For these conditions we find that
although the BCS coherence length (related to the $T=0$ pairing gap
$\Delta(0)$ by $\xi = \hbar^2 k_F/m\pi\Delta(0)$) is generally larger
than the size of the system, the finite size effects are negligible,
and we are thus allowed to relate the properties of the system to
those of an infinite homogeneous one with the same average
density. The calculations were carried out for the three different
interaction strengths $\lambda=0.2$, $\lambda=0.4$ and $\lambda=0.6$,
with the temperature fixed at the $T_c$ calculated accounting for the
direct contact interaction alone.

Before presenting the results we briefly review the well-known properties
of a homogeneous system.  Ignoring self-energy and induced
interaction, the (BCS) gap equation leads to the analytical result for the
critical temperature
\begin{equation}
k_B T_{c0}=\frac{e^{C}}{\pi}\frac{8}{e^2}\epsilon_F\;e^{-1/\lambda},
\end{equation}
$C\approx 0.577$ being Euler's constant.

Also in the weak coupling limit, however, the induced interaction must
be included in the calculation, since it leads to a constant decrease
in the critical temperature by a factor $(4e)^{1/3}\approx 2.2$
\cite{GORKOV}. The correct critical temperature for a uniform system
in the weak-coupling regime is therefore
\begin{equation}
k_B
T_c=\frac{e^{C}}{\pi}\left(\frac{2}{e}\right)^{7/3}\epsilon_F\;e^{-1/\lambda}.
\end{equation}
This result may be obtained using Eq. (\ref{Eq:gapfinalsigamma}) by replacing
in the expression for the effective interaction $\Pi_\rho$ and
$\Pi_{\sigma_i}$ with the unperturbed response function $\Pi^0_\rho$ (an
approximation valid in the weak-coupling limit). Within the present context,
one should also ignore the self-energy effects ($\gamma_\nu$, ${\rm
Re}(\Sigma_\nu)$ and $Z_\nu$), and perform an average at the Fermi surface of
the induced interaction \cite{GORKOVVIVERIT}.

As the interaction strength $\lambda$ is increased, both the
renomalization of the quasi-particle properties due to self-energy
effects and the collectivity of the modes become important.  One must
therefore keep all terms in Eq. (\ref{Eq:gapfinalsigamma}) and solve
the full eigenvalue equation.  For an infinite homogeneous system the
problem was considered in Ref. \cite{COMBESCOT}. The author used a
different set of approximations as compared to the ones we used.  He
considered the renormalization of the quasi-particle properties only
at the Fermi surface, where he performed an angular average in the
spirit of Gorkov and Melik-Barkhudarov \cite{GORKOV}. At the same time
the author kept the full $\omega$-dependence in the quantities,
i.e. he did not use the one pole approximation. Below we discuss the
comparison between our results and those of Ref. \cite{COMBESCOT} for
the critical temperature.

\begin{figure}[hh]
\vspace{1cm}
  \begin{center}
    \psfig{file=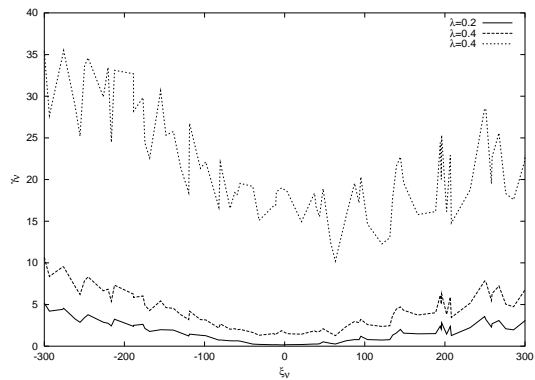,width=0.32\textwidth,angle=270}
    \caption{The figure shows the width $\gamma_\nu$ of the level
      $\nu$ as a function of the hartree energy $\xi_\nu$ for a
      spherically symmetric infinite square well. The three curves
      correspond to the interaction strengths $\lambda=0.2$ (solid
      line), $\lambda=0.4$ (long dashed line) and $\lambda=0.6$ (short
      dashed line). The energies are measured in the units of
      $\hbar^2/2m_aR^2$, with $R$ being the radius of the well.}
    \label{Fig:imunif}
  \end{center}
\end{figure}

Before doing so, we turn to the analysis of the self-energy effects
and the quasi-particle properties. In Fig.  \ref{Fig:imunif} we show
the width $\gamma_\nu$ of the level $\nu$, calculated using
Eq.~(\ref{sigma2}), as a function of the Hartree quasi-particle energy
$\xi_\nu$.  The three curves correspond to the values $\lambda=0.2$,
$\lambda=0.4$ and $\lambda=0.6$ of the interaction strength. The
figure shows a very strong dependence of $\gamma_\nu$ on the
interaction. In particular there is a very large increase in the level
width in going from $\lambda=0.4$ to $\lambda=0.6$. The large values of
$\gamma_\nu$ for $\lambda=0.6$ question the validity of the one pole
approximation for this or higher values of the interaction
parameter. As we shall see we reach similar conclusions also for the
harmonically trapped case.

\begin{figure}[hh]
\vspace{1cm}
  \begin{center}
    \psfig{file=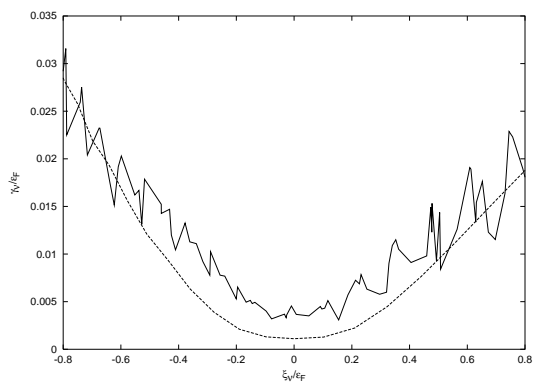,width=0.32\textwidth,angle=270}
    \caption{The curve in Fig. \ref{Fig:imunif} for $\lambda=0.4$
    (solid line) is compared with the one obtained for an infinite
    homogeneous system with the same interaction strength using the
    plane waves basis (dashed line). For the comparison the energies
    are here measured in units of the Fermi energy.}
    \label{Fig:imunifpaco}
  \end{center}
\end{figure}

In Fig. \ref{Fig:imunifpaco} we show the same curve as the one in
Fig. \ref{Fig:imunif} for $\lambda=0.4$ and compare it with that
obtained for an infinite system at the same interaction strength using
the plane waves basis. The calculation for the square well shows a
staggering due to the discrete shell structure considered. The overall
behavior is, nevertheless, very close to that of an infinite system.

\begin{figure}[hh]
\vspace{1cm}
  \begin{center}
    \psfig{file=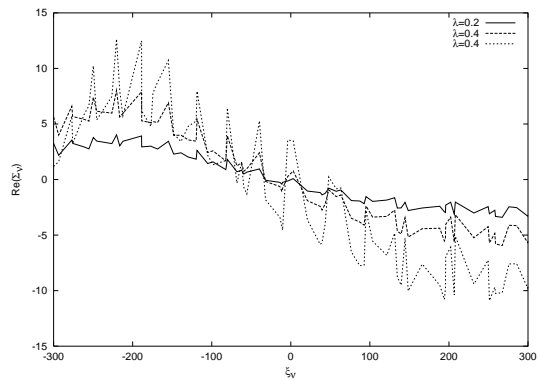,width=0.32\textwidth,angle=270}
    \caption{The figure shows the real part of the self-energy ${\rm
      Re}\Sigma_\nu(\epsilon_\nu)$ of the level $\nu$ as a function
      of the hartree quasi-particle energy $\xi_\nu$ for a spherically
      symmetric infinite square well. The three curves correspond to
      the same interaction strengths as in Fig. \ref{Fig:imunif} and
      the same line codes and energy units are used.}
    \label{Fig:reunif}
  \end{center}
\end{figure}

\begin{figure}[hh]
\vspace{1cm}
  \begin{center}
    \psfig{file=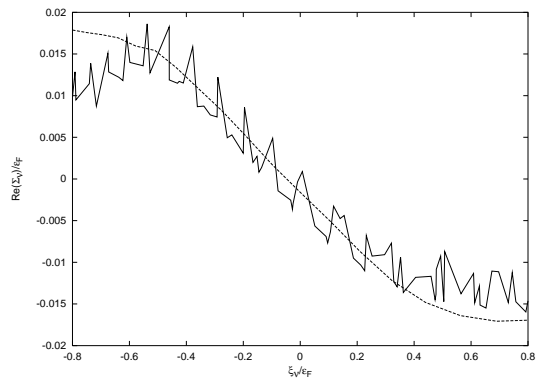,width=0.32\textwidth,angle=270}
    \caption{Here we compare the curve in Fig. \ref{Fig:reunif} for
    $\lambda=0.4$ (solid line) with the one calculated for an infinite
    homogeneous system using the plane waves basis (dashed line). The
    energies are measured in units of the Fermi energy.}
    \label{Fig:reunifpaco}
  \end{center}
\end{figure}

In Figs. \ref{Fig:reunif} and \ref{Fig:reunifpaco} we display the real part of
the self-energy. We observe that this quantity is almost symmetric about the
Fermi surface, where it changes sign. The net effect is an increase in the
density of states at the Fermi level (cf. e. g. \cite{MAHAUX} and
refs. therein).

\begin{figure}[hh]
\vspace{1cm} \begin{center}
\psfig{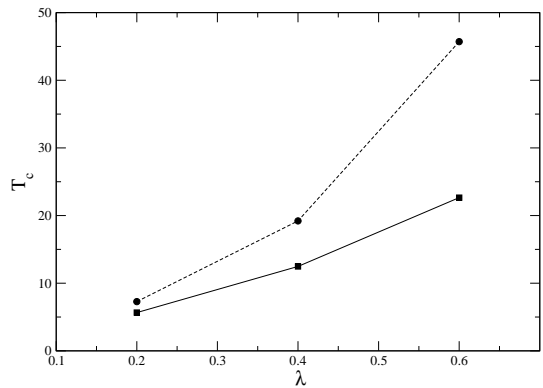} \caption{The critical
temperature $T_c$ is plotted against the interaction strength for the gas in
the spherically symmetric inifinite square well. The solid line is the result
of the calculation made including the full quasi-particle renormalization
properties. The dashed line corresponds to the calculation using the bare
interaction alone. $T_c$ is in units of $\hbar^2/2m_aR^2$.}
\label{Fig:Tcunif} \end{center}
\end{figure}

\begin{figure}[hh]
\vspace{1cm} \begin{center}
\psfig{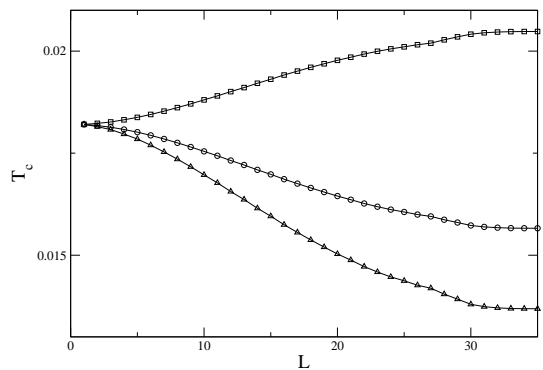} \caption{Critical
temperature of the gas in a spherical symmetric infinite square well, with
$\lambda=0.2$, as calculated from Eq. (\ref{Eq:gapfinalsigamma2}) including
only the effective interaction contribution, as a function of the maximum
multipolarity $L$ included in the correlation functions appearing in the
matrix elements in Eq. (\ref{Eq:matelv}). $L=1$ corresponds to the bare
interaction alone. Convergence is attained around $L=35$. $T_c$ is in units of
$\hbar^2/2m_aR^2$, $R$ being the radius of the well. Circles represent the
result of the calculation which includes the full induced interaction. Squares
are obtained accounting only for the exchange of density modes and triangles
only accounting for spin modes.}  \label{Fig:Tc_L_unif}
\end{center}
\end{figure}

  We now turn to the results for $T_c$. In Fig. \ref{Fig:Tcunif} we show the
value of $T_c$ as a function of $\lambda$. The dashed line corresponds to
calculations including the bare interaction alone, the solid line to
calculations including all quasi-particle renormalization effects. In Table
\ref{table:2} we report the results of the calculations when each single
renormalization contribution is included. The real part of the self-energy
gives rise to an increase in the level density around the Fermi surface (see
Fig. \ref{Fig:reunif}) and a consequent increase in $T_c$ due to the resulting
enhanced effective coupling strength.  Both $Z_\nu$ and $\gamma_\nu$ decrease
the value of $T_c$.  The former causes a weaker effective interaction between
quasi-particles, as can be deduced from Eq.~(\ref{Eq:gapfinalsigamma})
recalling that $Z_\nu\leq 1$.  Because $\gamma_\nu$ measures the energy range
over which the quasi-particle state is spread due to the coupling to
vibrations (lifetime), its presence effectively inhibits pairing between
quasi-particles \cite{MOREL}.  The net effect of the induced interaction is to
decrease the value of $T_c$, as can be seen by looking at the circles in
Fig. \ref{Fig:Tc_L_unif}. There we show the value of $T_c$ for $\lambda=0.2$, calculated
including only the contribution of the effective interaction, as a function of the
maximum multipolarity included in the calculation of the correlation functions
in Eq. (\ref{Eq:matelv}) (see the Appendix for details on the multipolar
expansion). If the overall effect of the combination of spin singlet (density) and
triplet (spin) modes is repulsive, isolating the two contributions shows that density
modes provide an attractive effective interaction (see squares in
Fig. \ref{Fig:Tc_L_unif}), and spin modes a repulsive one (triangles).

Our results indicate that $\lambda=0.2$ does not belong entirely to
the weak-coupling regime. If this were the case, in fact, inclusion of
the induced interaction would result in a reduction in $T_c$ by a
factor of the order of $2$ as predicted in Ref. \cite{GORKOV}. This is
however only true if $\Pi^r\simeq\Pi^0_\rho$. When $\lambda=0.2$ we
indeed find a reduction by factor $\simeq 2$ but only when
$\Pi^0_\rho$ is used instead of the full $\Pi^r$  in the matrix elements in
Eq. (\ref{Eq:matelv}). Using the full $\Pi^r$ we found a reduction only by a
factor $\sim 1.4$. We are lead to conclude that, at this density, the
collectivity of the modes still plays a significant role.

For what concerns the comparison with the results of
Ref. \cite{COMBESCOT}, we find a reduction in $T_c$ in the stronger
coupling regimes which is not at all as large as the one reported
there. We have carefully analyzed our set of approximations and we
know that they are less and less reliable as the interaction becomes
stronger. We are not aware of such a careful analysis for the
framework of Ref. \cite{COMBESCOT} which was, on the other hand, a
pioneering calculation in this field.  It is therefore not clear as
yet which is the best set of approximations to be used, and thus the
appropriate values for $T_c$ one should expect, as these appear to be
strongly model dependent.

\begin{table}
\begin{indented}
\item[]\begin{tabular}{|c|ccc|} \hline & \;\;\; $\lambda$=0.2\;\;\;
&\;\;\; $\lambda$=0.4\;\; \;&
\;\;\; $\lambda$=0.6\;\;\;\\
\hline \hline
$V^{eff}$ & 0.86  &  0.81 & 0.94\\
\hline
$g$+ Re$\Sigma^{ph}$ & 1.05 & 1.08 & 1.10\\
\hline
$g$+ Z & 0.89  & 0.84 & 0.80\\
\hline
$g$+ $\gamma$ & 0.97  & 0.92 & 0.70\\
\hline
$V^{eff}$+ Re$\Sigma^{ph}$ + $\gamma$ + Z & 0.77  & 0.65 & 0.49 \\
\hline
\end{tabular}
\caption{The table gives the ratio $T_c/T_c^{\rm bare}$ associated
with the indicated contribution for the case of a spherically
symmetric infinite square well. $T_c^{\rm bare}$ is calculated
including only $g$ and we find the following results: for
$\lambda=0.2$, $k_BT_c^{\rm bare}=7.3\hbar^2/2m_aR^2$, for
$\lambda=0.4$, $k_BT_c^{\rm bare}=19.2\hbar^2/2m_aR^2$ and for
$\lambda=0.6$, $k_BT_c^{\rm bare}=45.7\hbar^2/2m_aR^2$.}
\label{table:2}
\end{indented}
\end{table}

\subsection{Trapped system}

\label{sec:restrap}

Once again the calculations were carried out for the three different
interaction strengths $\lambda=0.2$, $\lambda=0.4$ and $\lambda=0.6$,
with the temperature fixed at the $T_c$ for the system interacting via
the direct contact interaction alone.

\begin{figure}[hh]
\vspace{1cm}
  \begin{center}
    \psfig{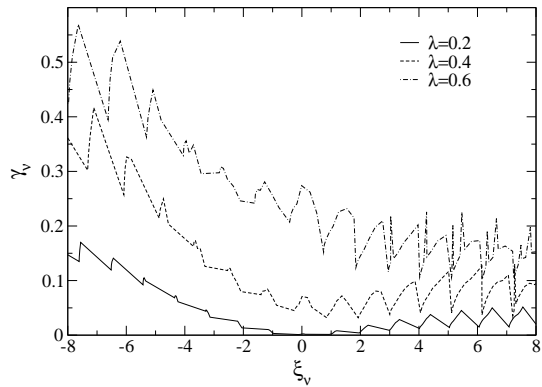}
    \caption{The width $\gamma_\nu$ of the level $\nu$ is plotted as a
      function of the hartree energy $\xi_\nu$ for a spherically
      symmetric harmonic trap. The three curves correspond to the
      interaction strengths $\lambda=0.2$ (solid line), $\lambda=0.4$
      (dashed line) and $\lambda=0.6$ (dot-dashed line). The energies
      are measured in units of $\hbar\omega_0$, with $\omega_0$
      being the oscillator frequency.}
    \label{Fig:im}
  \end{center}
\end{figure}

\begin{figure}[hh]
\vspace{1cm} \begin{center}
\psfig{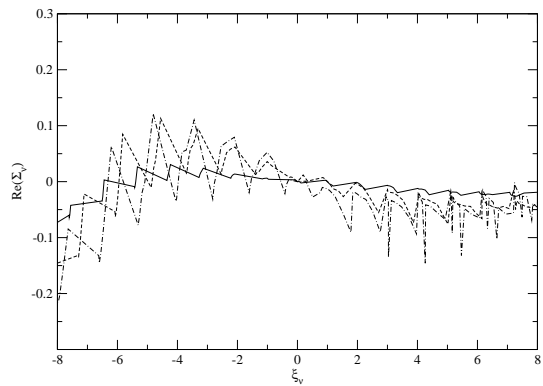}
\caption{The figure shows the real part of the self-energy ${\rm
Re}\Sigma_\nu^{ph}(\epsilon_\nu)$ of the level $\nu$ as a function of the
hartree quasi-particle energy $\xi_\nu$ for a spherically symmetric harmonic
trap. The three curves correspond to the same interaction strengths as in
Fig. \ref{Fig:im} and the same line codes and energy units are used.}
\label{Fig:re} \end{center}
\end{figure}

In Figs.  \ref{Fig:im} and \ref{Fig:re} we show the width
$\gamma_\nu$ and the real-part of the self energy $(\xi_\nu)$
respectively, as a function of $\xi_\nu$. Like for a uniform system
from Fig.~\ref{Fig:im} we see that the levels acquire an
increasingly larger width as the interaction strength is increased.
For $\lambda=0.2$ and $\lambda=0.4$ the width of the levels at the
Fermi surface ($\xi_\nu=0$) is small and the one pole approximation
is expected to be satisfactory.  For $\lambda=0.6$ one has
$\gamma_\nu \simeq 0.25 \hbar\omega_0$. Since the width is on the
one hand much larger than ${\rm Re}\Sigma_\nu^{ph} (\epsilon_\nu)$
and on the other almost of the order of the discrete level spacing
$\hbar\omega_0$, the one pole approximation is again probably not
very meaningful at this or higher values of $\lambda$, consistently
with the considerations made for a uniform system.

\begin{figure}[hh]
\vspace{1cm} \begin{center}
\psfig{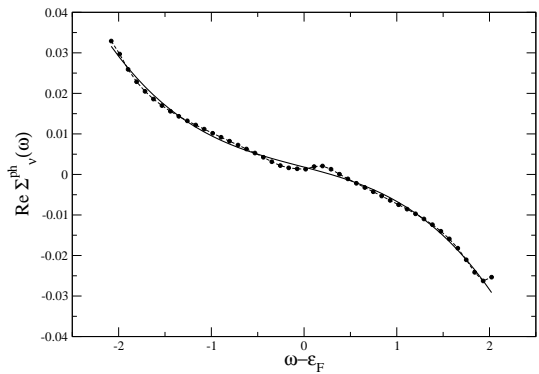}
\caption{Real part of the phonon self-energy ${\rm Re}\Sigma_\nu^{ph}
(\omega)$ relative to the state $\nu$ at the Fermi surface (dots), as a
function of $\omega-\epsilon_F$ being $\epsilon_F$ the Fermi energy, at
$\lambda=0.2$. The continuous line is the polynomial curve used to take the
derivative and calculate $Z_\nu$.}
\label{Fig:Z}
\end{center}
\end{figure}

The renormalization factor $Z_\nu$ requires special care when it is
evaluated for a finite sized system. In Fig.~\ref{Fig:Z} we show
${\rm Re}\Sigma_\nu^{ph} (\omega)$ for the level $\nu$ at the Fermi surface.
While the general trend of the function is definite and follows a
curve which resembles that of a uniform system, the discrete level
structure causes a staggering which makes the derivative
$\partial{\rm Re}\Sigma_\nu^{ph} (\omega)/\partial\omega$ not well defined.
We have therefore interpolated ${\rm Re}\Sigma_\nu^{ph} (\omega)$ with a
polynomial before taking the derivative.

For $\lambda=0.2$ and $\lambda=0.4$ the value of $Z_\nu$ about the Fermi
surface has shown a very weak dependence on the degree of the polynomial
chosen for the interpolation (cf. also Ref. \cite{MAHAUX}).  It is remarkable
that the function ${\rm Re}\Sigma_\nu^{ph} (\omega)$ also has a very weak
dependence on $\nu$. The one reproduced in Fig.~\ref{Fig:Z} very closely
resembles those evaluated for different values of $\nu$. Since the staggering
increases in size as one moves away from the Fermi surface, both the
interpolation and the derivative become more and more ill-defined. However the
contribution of these levels to the gap equation is negligible, for this
reason the procedure used has demonstrated to be solid and so has the
calculated value of $T_c$.  For $\lambda=0.6$ on the other hand the staggering
is large also at the Fermi surface.  This once again points at the fact that
the approximation used may not be entirely applicable for this interaction
strength.

\begin{figure}[hh]
\vspace{1cm}
  \begin{center}
    \psfig{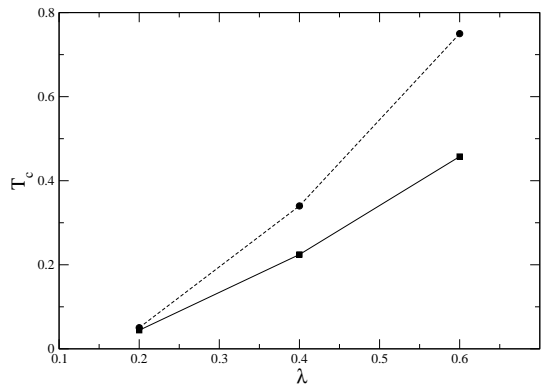}
    \caption{Critical temperature of a gas of $1000$ particles in a
    spherically symmetric harmonic oscillator. $T_c$ is in units of
    $\hbar\omega_0$. The dashed curve corresponds to $T_c$ calculated
    using the bare interaction alone, the solid one is obtained
    including all quasi-particle renormalization effects.}
    \label{Fig:Tc}
  \end{center}
\end{figure}

We now turn to discuss the effects on $T_c$.
In Fig.~\ref{Fig:Tc} we show the results for $T_c$ as a function of
$\lambda$. The two curves correspond to $T_c$ as evaluated using the
bare interaction alone (dashed curve), and including all
quasi-particle renormalization effects (solid curve).

\underline{General features} - In Table \ref{Tab:1} we report the
results of the calculations when the specific effects of each
renormalized quasi-particle property is isolated. The effects of the
self-energy renormalizations are analogous to those we dealt with in
the uniform case. The real part ${\rm Re}\Sigma^{ph}$ causes an
increase in the critical temperature since, as before, it leads to an
effective increase of the density of levels at the Fermi energy. On
the other hand, both $Z_\nu$ and $\gamma_\nu$ depress the value of
$T_c$, the former by causing a weaker effective interaction between
quasi-particles and the latter by inhibiting pairing
between quasi-particles \cite{MOREL}.

\underline{Specific features} - In the weak coupling regime
$\lambda=0.2$ the critical temperature is hardly affected by the
renormalization of the quasi-particle properties. While this is
expected as far as $\gamma_\nu$, $Z_\nu$ and ${\rm
Re}\Sigma_\nu^{ph}$ are concerned, it is instead surprising as
regards the effect of the induced interaction. For a uniform system at
$\lambda=0.2$ we have seen that, if the induced interaction is
calculated using only $\Pi_\rho^{0}$ one finds a decrease in $T_c$ by
a factor of the order of 2. This effect is reduced, but still
present, when the collectivity of the modes is accounted for. Almost
no reduction is obtained instead for a trapped gas, whether one uses
the lowest-order correlation function $\Pi_\rho^{0}$ or one accounts
for the full collectivity of the modes. The absence of the reduction
descends here from the discrete level structure of the harmonic
trap. This can be understood by comparing the strength functions of a
uniform system with those of a harmonically trapped one for the same
interaction strength, see Figs. \ref{Fig:resptrapL2_02},
\ref{Fig:resptrapL12_02}, \ref{Fig:respunifL2_02} and
\ref{Fig:respunifL12_02}.  The largest contributions to the sum in
Eq.~(\ref{Eq:gapfinalsigamma}) comes from the condition
$\epsilon_\nu=\epsilon_{\nu'}=\omega=0$, which corresponds to
particles at the Fermi surface and to the exchange of a phonon with
zero frequency. The discrete level structure causes the values of the
strength functions to be negligible near $\omega=0$, and consequently
the effect of the phonon-induced interaction to be small.
As the strength of the interaction increases ($\lambda=0.4$ and
$\lambda=0.6$) the renormalization of the quasi-particle properties
has a larger and larger effect on $T_c$. All the terms are
increasingly more important but it is the quasi-particle width that
dominates over the others, as can be seen in Table \ref{Tab:1}. We
recall however that the case $\lambda=0.6$ is beyond the validity of
the one pole approximation.

\begin{center}
\begin{table}
\begin{indented}
\item\begin{tabular}{|c|ccc|} \hline & \;\;\; $\lambda$=0.2\;\;\;
&\;\;\; $\lambda$=0.4\;\; \;&
\;\;\; $\lambda$=0.6\;\;\;\\
\hline \hline
$V^{eff}$ & 0.92  &  0.86 & 0.90\\
\hline
$g$+ Re$\Sigma^{ph}$ & 1.01 & 1.03 & 1.03\\
\hline
$g$+ Z & 0.98  & 0.90 & 0.93\\
\hline
$g$+ $\gamma$ & 0.97  & 0.85 & 0.76\\
\hline
$V^{eff}$+ Re$\Sigma^{ph}$ + $\gamma$ + Z & 0.88  & 0.66 & 0.61 \\
\hline
\end{tabular}
\caption{The table gives the ratio $T_c/T_c^{\rm bare}$ associated with the
indicated contribution for the case of a spherically symmetric harmonic
potential. $T_c^{\rm bare}$ is calculated including only $g$ and we find the
following results: for $\lambda=0.2$, $k_BT_c^{\rm
bare}=5.03\times10^{-2}\hbar\omega_0$, for $\lambda=0.4$, $k_BT_c^{\rm
bare}=0.34\hbar\omega_0$ and for $\lambda=0.6$, $k_BT_c^{\rm
bare}=0.75\hbar\omega_0$.}
\label{Tab:1}
\end{indented}
\end{table}
\end{center}

\begin{figure}[hh]
\vspace{1cm} \begin{center}
\psfig{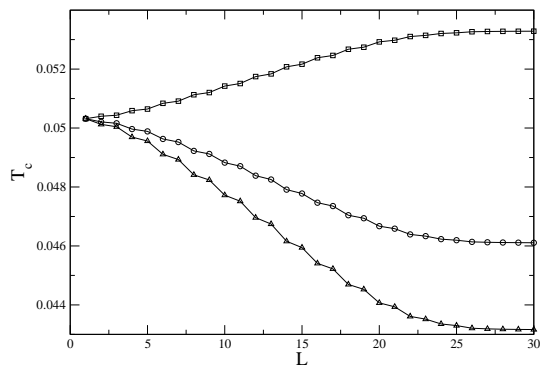} \caption{Critical
temperature of the harmonically trapped gas with $\lambda=0.2$, as calculated
from Eq. (\ref{Eq:gapfinalsigamma2}) including only the effective interaction
contribution, as a function of the maximum multipolarity $L$ included in the
correlation functions appearing in the matrix elements in
Eq. (\ref{Eq:matelv}). $L=1$ corresponds to the bare interaction
alone. Convergence is reached around $L=27$. $T_c$ is in units
$\hbar\omega_0$, with $\omega_0$ being the trap frequency. Circles represent
the result of the calculation which includes the full induced
interaction. Squares are obtained accounting only for the exchange of density
modes and triangles only accounting for spin modes. } \label{Fig:tctrapL02}
\end{center}
\end{figure}

\section{Conclusions}
\label{sec:conclusions}

In this paper we have calculated the renormalization of the
quasi-particle properties in an interacting Fermi gas due to the
emission and reabsorption of density and spin fluctuations of the
system.  In particular we have looked, within the one pole
approximation, at  self-energy ($\Sigma$) and screened
interaction ($V^{eff}$) effects. The real part of $\Sigma$ leads to a
shift in the energy of the quasi-particles and to a reduction in the
strength of the quasi-particle peak of the associated propagator (by a
factor $Z\leq 1$). At the same time a non-vanishing imaginary part of
$\Sigma$ results in a non-vanishing width $\gamma$ of the
levels, or equivalently in a finite lifetime of the quasi-particles.
The exchange of collective modes between quasi-particles leads to an
effective interaction which has an overall repulsive character. This
is due to the dominance of the spin triplet (repulsive) modes over
spin singlet (attractive) modes, and leads to a screening of the bare
attractive interaction.

Finally we have calculated the effect of these properties on the
critical temperature to the superfluid state. The calculations were
carried out numerically for the two illustrative cases of an isotropic
infinite square well and for a harmonic trap. In both cases the
calculations were carried out for a sytem of 1000 particles.  We have
included selectively one or the other of the renormalized
quasi-particle properties and found that these have different
consequences of $T_c$. When all of them are accounted for, the overall
effect is a reduction of $T_c$ which depends both on the strength of
the interaction and on the geometry of the system.  The results are
reported in Sec. \ref{sec:results}. As the strength of the interaction
increases, although the one pole approximation becomes less and less
reliable, our calculations clearly indicate that the effect of
fluctuations becomes more and more important and modifies the critical
temperature, as compared to the one obtained using the bare
quasi-particle properties, in an important way.  This result is
very relevant for current experiments with atomic Fermi gases. In
particular it shows that the standard theories of the crossover from
BCS to BEC, which are currently used to predict the value of the
critical temperature also in the strongly interacting regime may be
wrong by a significant amount as the interaction becomes stronger.  Our
approach is well suited to calculate the first corrections to the
quasi-particle properties and to the critical temperature which arise
when one moves towards the strongly interacting regime originating
from the BCS weak coupling side.

\appendix
\section*{Appendix}
\setcounter{section}{1}

\subsection{The correlation functions}

\label{appcorr}

Define:
\begin{eqnarray}
\nonumber \Pi_{\alpha\beta}(x,x')= -e^{\beta\Omega} {\rm
Tr}\big\{\hat\rho \;T_\tau[ \tilde n_\alpha(x)\tilde n_\beta(x')]\big\}
\end{eqnarray}
where $\tilde n_\sigma(x)=\hat n_\sigma(x)
-\langle\cpsi{\sigma}{\vr}\dpsi{\sigma}{\vr}\rangle$, with $\hat
n_\sigma(x)=\cpsi{\sigma}{x}\dpsi{\sigma}{x}$ and
$x\equiv(\vr,\tau)$. The indices $\alpha$, $\beta$ and $\sigma$ can
assume the two values $\up$ and $\down$, the operator $\hat\rho$ is
given by $\hat\rho=\exp[\hat H-\mu\hat N]$, with $\hat N=\sum_\sigma
\int d^3r\; \cpsi {\sigma}{\vr}\dpsi {\sigma}{\vr}$ and $T_\tau$ is
the ordering operator relative to the imaginary time $\tau$. Finally
$\exp(-\beta\Omega)$ is the grand partition function of the system,
and $\Omega$ the thermodynamic potential.

\underline{Density correlation function} - Let us now introduce the density
correlation function with the definition:
\begin{eqnarray}
\Pi_{\rho}(x,x')=-e^{\beta\Omega} {\rm Tr}\big\{\hat\rho \;T_\tau[
\tilde n(x)\tilde n(x')]\big\}
\end{eqnarray}
where $\tilde n=\tilde n_\up+\tilde n_\down$.
This correlation function is related to the previously defined
$\Pi_{\alpha\beta}$ as follows:
\begin{eqnarray}
\Pi_\rho(x,x')=2\Pi_{\up\up}(x,x')+2\Pi_{\up\down}(x,x'),
\end{eqnarray}
where we have used $\Pi_{\up\up}(x,x')=\Pi_{\down\down}(x,x')$
and $\Pi_{\up\down}(x,x')=\Pi_{\down\up}(x,x')$.

\underline{Spin correlation function} - The spin correlation function
has components along the three directions $x$, $y$ and $z$, defined as
follows:
\begin{eqnarray}
\Pi_{\sigma_i}(x,x')-e^{\beta\Omega} {\rm Tr}\big\{\hat\rho
\;T_\tau[\hat\sigma_i(x)\hat\sigma_i(x')]\big\}
\end{eqnarray}
where $\hat\sigma_i=\sum_{\alpha\beta}\cpsi{\alpha}{x}
(\sigma_i)_{\alpha\beta}\dpsi{\beta}{x}$, and $(\sigma_i)_{\alpha\beta}$
being the Pauli spin matrix element.
The spin correlation function in the $z$ direction
is related to $\Pi_{\alpha\beta}$ by:
\begin{equation}
\Pi_{\sigma_z}(x,x')=2\Pi_{\up\up}(x,x')
-2\Pi_{\up\down}(x,x').
\end{equation}
Finally since the choice of the orientation of the axes is arbitrary
we must also have
$\Pi_{\sigma_x}(x,x')=\Pi_{\sigma_y}(x,x')=\Pi_{\sigma_z}(x,x')$.  The
$\Pi_{\alpha\beta}$ therefore contains all the information needed.

$\Pi_{\alpha\beta}$ admits the Lehmann representation:
\begin{eqnarray}
\fl \nonumber &\Pi_{\alpha\beta}(\vr,\vr',i\omega_n)=e^{\beta\Omega}
(e^{-\beta E_n}-e^{-\beta E_m}) \sum_{n,m\neq0}\displaystyle
\frac{\langle \Phi_m| \tilde n_\alpha(\vr)|\Phi_n\rangle\langle
\Phi_n| \tilde n_\beta(\vr') |\Phi_m\rangle} {i\omega_m-(E_n-E_m)}
\label{Eq:bigpi}
\end{eqnarray}
where the complete basis of eigenstates $|\Phi_n\rangle$ introduced above is
in principle the exact basis of excited states of the system.  Below
we shall assume specific approximations for the $|\Phi_n\rangle$'s.

\subsection{The RPA approximation}

To evaluate $\Pi_{\alpha\beta}$ one has to introduce specific
approximations.
The lowest order approximation is the Hartree-Fock particle-hole one,
in which case:
\begin{equation}
\Pi_{\alpha\beta}^{hf}(\vr,\vr',i\omega_n)=\left(
\begin{array}{cc}
\chi_0(\vr,\vr',i\omega_n) & 0 \\
0 & \chi_0(\vr,\vr',i\omega_n)
\end{array}
\right).
\end{equation}
As is well known a better approximation for $\Pi_{\alpha\beta}$ is
the RPA self-consistent one.
In this other case $\Pi_{\alpha\beta}$ is assumed to
satisfy the equation (for each given $i\omega_n$):
\begin{eqnarray}
\nonumber
&\left(
\begin{array}{cc}
\Pi_{\up\up}^{r}(\vr,\vr') & \Pi_{\up\down}^{r}(\vr,\vr') \\
\Pi_{\down\up}^{r}(\vr,\vr') & \Pi_{\down\down}^{r}(\vr,\vr')
\end{array}
\right)\left(\begin{array}{cc}
\chi_0(\vr,\vr') & 0 \\
0 & \chi_0(\vr,\vr')
\end{array}
\right)\\
\nonumber
&+\int d^3r_1 d^3 r_2\;
\left(\begin{array}{cc}
\chi_0(\vr,\vr_1) & 0 \\
0 & \chi_0(\vr,\vr_1)
\end{array}
\right)\\
\nonumber
&
\left(\begin{array}{cc}
0 & g\delta(\vr_1-\vr_2)\\
g\delta(\vr_1,\vr_2) & 0
\end{array}
\right)\left(
\begin{array}{cc}
\Pi_{\up\up}^{r}(\vr_2,\vr') & \Pi_{\up\down}^{r}(\vr_2,\vr') \\
\Pi_{\down\up}^{r}(\vr_2,\vr') & \Pi_{\down\down}^{r}(\vr_2,\vr')
\end{array}
\right).
\end{eqnarray}
In compact notation one can write
$\Pi_{\up\up}^{r}=\Pi_{\down\down}^{r}=\chi_0+g^2\chi_0^3+...$
and
$\Pi_{\up\down}^{r}=\Pi_{\down\up}^{r}=g\chi_0^2+g^3\chi_0^4+...$
Consequently $\Pi_\rho^{r}=2\chi_0/[1-g\chi_0]$ and
$\Pi_{\sigma_z}^{r}=2\chi_0/[1+g\chi_0]$.

\subsection{Multipolar expansion}

The spherical symmetry of our system allows
expansion in multipoles:
\begin{equation}
\Pi_{\alpha\beta}({\bf r},{\bf r}')=\sum_{LM}
[\Pi_{\alpha\beta}(r,r')]_L Y_{LM}(\Omega)
Y_{LM}^*(\Omega')
\label{multipol}
\end{equation}
where $\Omega$ and $\Omega'$ are the angular coordinates of $\vr$
and $\vr'$ respectively.

\subsection{Strength function and sum rules}

\begin{figure}[hh]
\vspace{1cm}
  \begin{center}
    \psfig{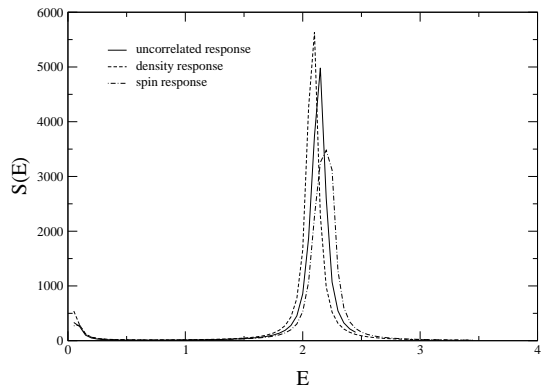}
    \caption{The figure shows the strength functions for density and
spin modes with multipolarity $L=2$, for a system of 1000 particles in
a sperically symmetric harmonic trap and interaction strength
$\lambda=0.2$. The solid line shows the strength functions calculated
with the Hartree-Fock correlation function $\Pi^{hf}$, in this case
density and spin response coincide. In the RPA the two responses and
distinct and are given by the dashed line and dot-dashed line
respectively.  The energy on the $x$-axis is measured in units of the
oscillator level spacing $\hbar\omega_0$.}
  \label{Fig:resptrapL2_02}
  \end{center}
\end{figure}
\begin{figure}[hh]
\vspace{1cm}
  \begin{center}
    \psfig{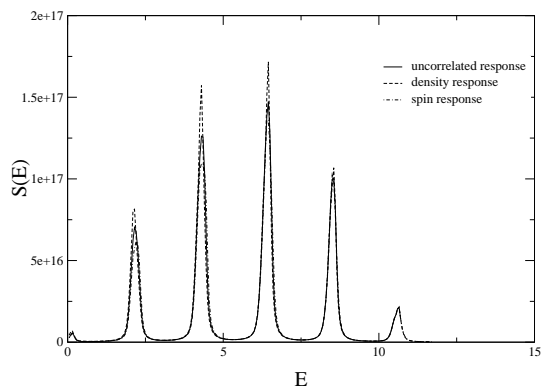}
    \caption{Same as Fig. \ref{Fig:resptrapL2_02} for the $L=12$ mode.}
    \label{Fig:resptrapL12_02}
  \end{center}
\end{figure}
\begin{figure}[hh]
\vspace{1cm}
  \begin{center}
    \psfig{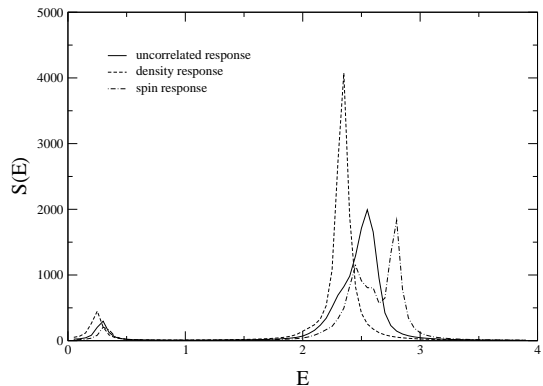}
    \caption{Same as Fig. \ref{Fig:resptrapL2_02} for the interaction
    $\lambda=0.4$.}
    \label{Fig:resptrapL2_06}
  \end{center}
\end{figure}
\begin{figure}[hh]
\vspace{1cm}
  \begin{center}
    \psfig{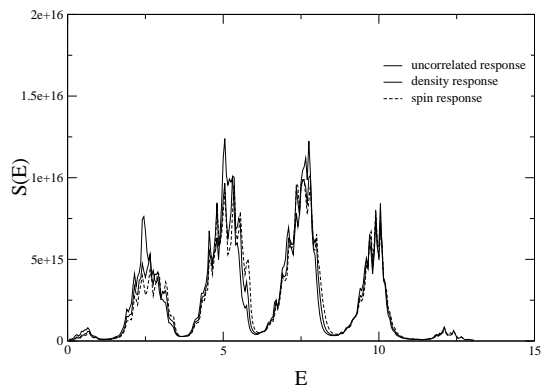}
    \caption{Same as Fig. \ref{Fig:resptrapL2_02} for the $L=12$ mode
    and interaction $\lambda=0.4$.}
    \label{Fig:resptrapL12_06}
  \end{center}
\end{figure}
\begin{figure}[hh]
\vspace{1cm}
  \begin{center}
    \psfig{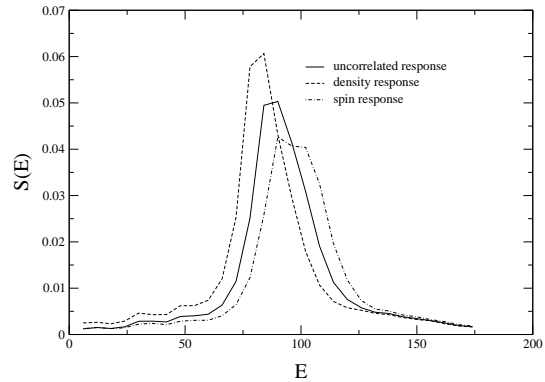}
    \caption{Strength function of the $L=2$ mode for a fermi gas in an
      isotropic infinite square well of radius $R$. The energy is in units
      of $\hbar^2/2mR^2$. The line codes are the same as for the previous
      figures.}
    \label{Fig:respunifL2_02}
  \end{center}
\end{figure}
\begin{figure}[hh]
\vspace{1cm}
  \begin{center}
    \psfig{file=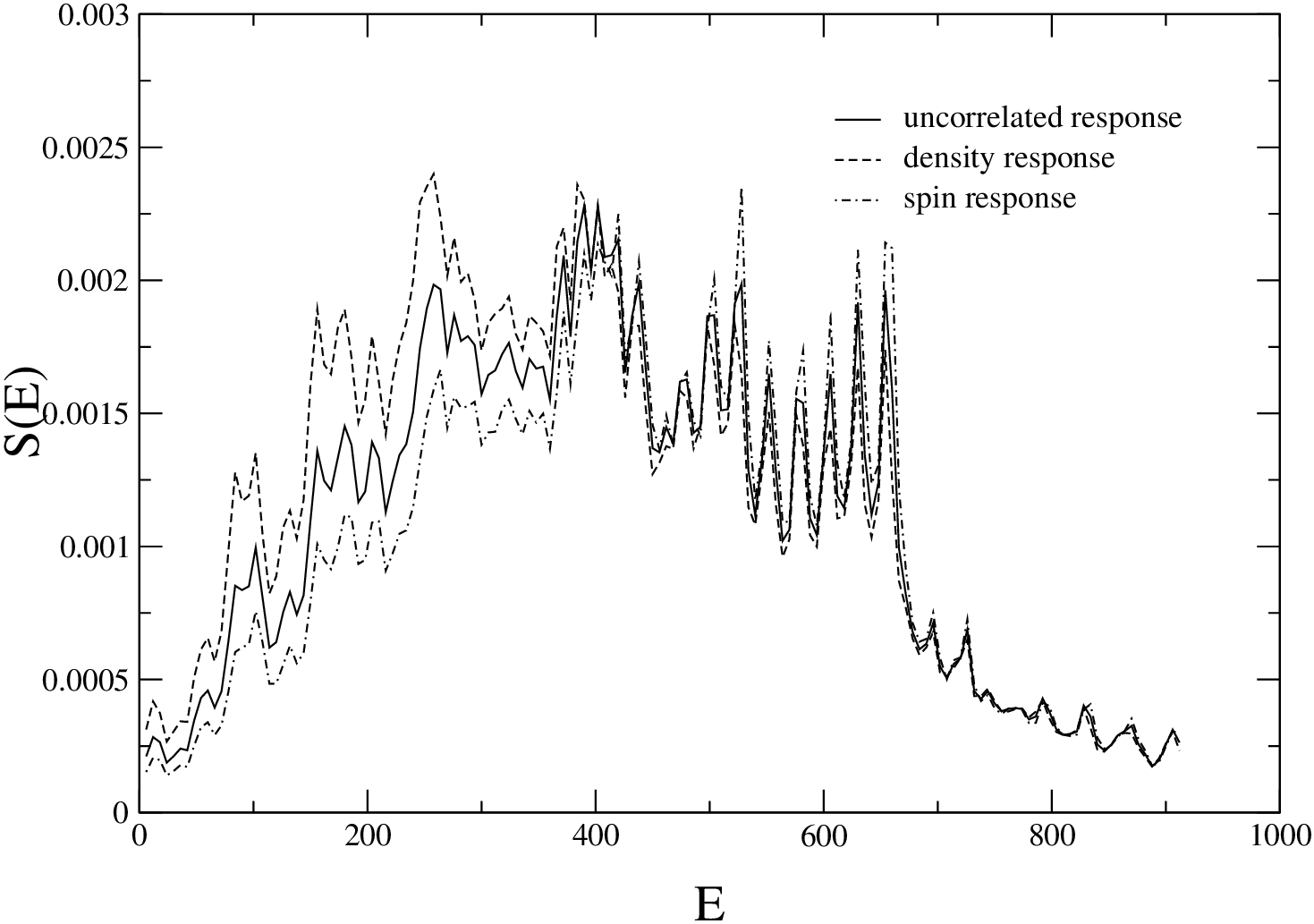,width=0.45\textwidth,angle=0}
    \caption{Same as Fig. \ref{Fig:respunifL2_02} for $L=12$.}
    \label{Fig:respunifL12_02}
  \end{center}
\end{figure}

An essential tool to study the collective modes is the strength
function.  This is defined as
\begin{eqnarray}
\nonumber
&S(F,E)=e^{\beta\Omega} \displaystyle\sum_{nm}(e^{-\beta E_n}-e^{-\beta E_m})\\
&\times|\langle m|\hat{F}|n \rangle|^2 \delta(E-E_{nm}),
\end{eqnarray}
where $\hat{F}=\sum_{\sigma}\int d^3r F_{\sigma}({\bf r})
\hat{\psi}^{\dagger}_{\sigma}({\bf r})\hat{\psi}_{\sigma}({\bf r})$
is the operator which corresponds to external fields
$F_{\uparrow}({\bf r})$ acting on the $\uparrow$ particles
and $F_{\downarrow}({\bf r})$ acting on the $\downarrow$ ones.
$E_{nm}$ stands for $E_n-E_m$.

The strength function is related to the retarded density correlation
function by
\begin{eqnarray}
\nonumber & S(F,E)=-\displaystyle\frac{1}{\pi}\sum_{\sigma\sigma'}
\int d^3r\,d^3r'\, F^*_{\sigma}({\bf r}) F_{\sigma'}({\bf r}') {\rm
Im}\{\Pi^0_{\sigma\sigma'} ({\bf r},{\bf r}',E)\}. \label{sfpi}
\end{eqnarray}
Density modes (i.e. with spin $s=0$) are excited by external fields such that
$F_{\uparrow}({\bf r})=F_{\downarrow}({\bf r})$, while spin modes
(with $s=1,s_z=0$) are excited by fields with
$F_{\uparrow}({\bf r})=-F_{\downarrow}({\bf r})$.

In the particular case of a density mode of given multipolarity $L$,
for which $F_{\uparrow}({\bf r})=F_{\downarrow}({\bf r})=r^L
Y_{LM}(\Omega)$, upon use of Eq.~(\ref{multipol}), this becomes
\begin{eqnarray}
\nonumber &S_L(E)=-\displaystyle\frac{1}{\pi}\sum_{\sigma\sigma'}
\int dr\,dr'\, r^{L+2}{r'}^{L+2} {\rm
Im}\{[\Pi^0_{\sigma\sigma'}(r,r',E)]_L\}. \label{sumrl}
\end{eqnarray}

Also useful is the energy weighted sum-rule \cite{BERBRO}
\begin{equation}
\int dE \; S(F,E)\, E=\frac{\hbar^2}{2m}\sum_{\sigma}\int d^3r\left|\nabla
F_{\sigma}({\bf r})\right|^2 n_{\sigma}(r)
\end{equation}
When $F(\vr)=r^L Y_{LM}(\Omega)$ it takes the simple form
\begin{eqnarray}
\nonumber \int
dE\;S_L(E)E&=-\displaystyle\frac{1}{\pi}\sum_{\sigma\sigma'} \int
dr\,dr'\, r^{L+2}{r'}^{L+2}
\times{\rm Im}\{[\Pi^0_{\sigma\sigma'}(r,r',E)]_L\}\\
&=\displaystyle\frac{\hbar^2}{2m}L(2L+1)\int_0^{\infty}
dr\;r^{2L}n_{\sigma}(r). \label{sumrl2}
\end{eqnarray}
and can be used as a check for the numerical calculations.

\end{fmffile}

\begin{thebibliography}{100}
\bibitem{note2} The Random Phase Approximation is in general a
convenient tool to describe the oscillations of the mean field around
the equilibrium shape, see \cite{BERBRO}.
\end{thebibliography}
\end{document}